\def \beq {\begin{equation}}
\def \eeq {\end{equation}}
\def \beqa {\begin{eqnarray}}
\def \eeqa {\end{eqnarray}}
\def \lf {\left}
\def \ri {\right}
\def \sw {\mbox{$\sigma\!-\!\omega$} }
\def \ed {{\cal E}}
\def \ep {\varepsilon}
\def \scdot {\!\cdot\!}
\newcommand{\ur}[1]{^{\rm #1}}
\newcommand{\dr}[1]{_{\rm #1}}
\newcommand{\req}[1]{(\ref{#1})}
\newcommand{\mcl}[1]{\multicolumn{#1}}
\newcommand{\ANP}[1]{Adv.\ Nucl.\ Phys.\ #1}
\newcommand{\AP}[1]{Ann.\ Phys.\ (N.Y.) #1}
\newcommand{\JPG}[1]{J. Phys.\  G #1}
\newcommand{\NPA}[1]{Nucl.\ Phys.\ A #1}
\newcommand{\PLB}[1]{Phys.\ Lett.\ B #1}

\newcommand{\PRC}[1]{Phys.\ Rev.\ C #1}
\newcommand{\PRL}[1]{Phys.\ Rev.\ Lett.\ #1}
\newcommand{\PRp}[1]{Phys.\ Reports #1}
\newcommand{\RPP}[1]{Rep.\ Prog.\ Phys.\ #1}
\newcommand{\ZPA}[1]{Z. Phys.\  A #1}
\newcommand{\ZPD}[1]{Z. Phys.\  D #1}
\documentstyle[12pt]{article}

\makeatletter
 \@addtoreset{equation}{section}
 
\makeatother
\makeatletter
\newcount\@tempcntc
\def\@citex[#1]#2{\if@filesw\immediate\write%
                  \@auxout{\string\citation{#2}}\fi
  \@tempcnta\z@\@tempcntb\m@ne\def\@citea{}\@cite{\@for\@citeb:=#2\do
    {\@ifundefined
       {b@\@citeb}{\@citeo\@tempcntb\m@ne\@citea%
                   \def\@citea{,}{\bf ?}\@warning
       {Citation `\@citeb' on page \thepage \space undefined}}%
    {\setbox\z@\hbox{\global\@tempcntc0\csname b@\@citeb%
                     \endcsname\relax}%
     \ifnum\@tempcntc=\z@ \@citeo\@tempcntb\m@ne
       \@citea\def\@citea{,}\hbox{\csname b@\@citeb\endcsname}%
     \else
      \advance\@tempcntb\@ne
      \ifnum\@tempcntb=\@tempcntc
      \else\advance\@tempcntb\m@ne\@citeo
      \@tempcnta\@tempcntc\@tempcntb\@tempcntc\fi\fi}}\@citeo}{#1}}
\def\@citeo{\ifnum\@tempcnta>\@tempcntb\else\@citea\def\@citea{,}%
  \ifnum\@tempcnta=\@tempcntb\the\@tempcnta\else
   {\advance\@tempcnta\@ne\ifnum\@tempcnta=\@tempcntb%
     \else \def\@citea{--}\fi
    \advance\@tempcnta\m@ne\the\@tempcnta\@citea\the\@tempcntb}\fi\fi}
\makeatother
\begin{document}
% >>>>>>>>>>>>>>>>>>>>>>>>>>>>>>>>>>>>>>>>>>>>>>>>>>>>>>>>>>>>>>>>>>>>
% TITLE AND AUTHORS.
%
\vspace*{-1.5cm}
\begin{center}
{\Large\bf Asymmetric semi-infinite nuclear matter:   \\[2mm]
           surface and curvature properties in        \\[5mm]
           relativistic and non-relativistic models}
\\[2.0cm]
M. Centelles, M. Del Estal, X. Vi\~nas  \\[2mm]
{\it Departament d'Estructura i Constituents de la Mat\`eria,
     Facultat de F\'{\i}sica,
\\
     Universitat de Barcelona,
     Diagonal {\sl 647}, E-{\sl 08028} Barcelona, Spain}
\end{center}
% >>>>>>>>>>>>>>>>>>>>>>>>>>>>>>>>>>>>>>>>>>>>>>>>>>>>>>>>>>>>>>>>>>>>
% ABSTRACT, PACS.
%
\vspace*{1.5cm}
\begin{abstract}
Surface and curvature properties of asymmetric semi-infinite nuclear
matter are studied to beyond the proton drip. Using the semiclassical
extended Thomas--Fermi method, the calculations are performed in the
non-relativistic and relativistic mean field theories (Skyrme forces
and non-linear \sw parametrizations). First we discuss the bulk
equilibrium between the nuclear and drip phases. Next we analyze the
asymmetric surface as a function of the bulk neutron excess. We
examine local quantities related to the density profiles and, for two
definitions of the bulk reference energy, the surface and curvature
energy coefficients. The calculation of the curvature energy is
carefully treated. The sensitivity of the nuclear surface to the
relativistic effects is investigated. Mass formulae useful for
arbitrary neutron excess are discussed, and their limit at small
asymmetries is compared with the liquid droplet model mass formula.
\end{abstract}

\mbox{}


\pagebreak
% >>>>>>>>>>>>>>>>>>>>>>>>>>>>>>>>>>>>>>>>>>>>>>>>>>>>>>>>>>>>>>>>>>>>
% INTRO.
%
\section{Introduction}
\hspace*{\parindent}
In several problems of nuclear physics and astrophysics the surface
and curvature properties of nuclei play a crucial role. This is the
case, for instance, of barrier heights and saddle-point configurations
in nuclear fission, or of fragment distributions in heavy-ion
collisions. In astrophysical applications they are important for
equilibrium sizes, electron capture rates and level densities used in
describing neutron stars and supernovae. However, the proton fraction
of the nuclei involved in these problems is rather different. For
terrestrial nuclei the proton fraction is around 0.4--0.5, whereas the
astrophysical problems demand a smaller proton concentration and,
furthermore, to consider nuclei surrounded by an external gas of drip
neutrons. Therefore, there is a strong motivation for studying the
surface and curvature properties of nuclei in the whole range of
asymmetries, between nuclear and neutron matter.

Within a context related to the liquid droplet model (LDM) and the
leptodermous expansion of a finite nucleus \cite{Mye69}, the surface
and curvature energy coefficients can be extracted from semi-infinite
nuclear matter. This schematic geometry avoids undesired shell,
Coulomb and finite-size effects, which makes it very appropriate to
study surface properties. Many of the calculations have been carried
out using the semiclassical extended Thomas--Fermi (ETF) method
\cite{Gra79,Rin80,Bra85} together with Skyrme forces \cite{Vau72}. For
the case of symmetric semi-infinite matter, the reader can consult
e.g.\ Refs.\ \cite{Bra85,Tre86,Sto88,Dur93}. Although the surface
energy can be calculated in a fully quantal way, only one part of the
curvature energy can be obtained from a quantum-mechanical calculation
\cite{Dur93,Cen96}. This is an important reason to compute the surface
and curvature coefficients by means of a semiclassical approach.

As far as asymmetric semi-infinite nuclear matter is concerned, we may
point out two publications of particular relevance for the present
work. The surface energy of multicomponent systems up to the proton
drip was investigated by Myers {\it et al.}\ \cite{Mye85} in the
Thomas--Fermi (TF) approximation with the Seyler--Blanchard
interaction. Almost at the same time, Kolehmainen {\it et al.}\
\cite{Kol85} performed an analysis of the surface and curvature
effects of neutron-rich nuclei, including drip neutrons, with the ETF
method and Skyrme forces.

In recent years the relativistic treatment of the nuclear many-body
problem has been a subject of growing interest
\cite{Ser86,Cel86,Rei89,Ser92}. The relativistic theory, already in
the mean field (Hartree) approximation, automatically includes the
spin-orbit force and the finite range and density dependence of the
nuclear interaction. All these effects are very important for the
surface properties of a nuclear system \cite{Fri75}. The
phenomenological \sw model of Walecka and its extensions
\cite{Ser86,Bog77} have become very popular in relativistic nuclear
calculations. They have been widely employed in the Hartree approach
for describing ground-state properties of nuclei
\cite{Hor81,Bou84,Rei86,Gam90} as well as symmetric semi-infinite
nuclear matter \cite{Bog77,Hof89,Von94a}.

Until recently the semiclassical approach to the relativistic mean
field theory had been settled at the TF level only
\cite{Ser86,Bog77,Bog77b,Sto91,Sha91}. However, in the last years some
amount of work has been addressed to develop the relativistic extended
Thomas--Fermi (RETF) method \cite{Cen90,Cen93b,Spe92,Von92b}, that
includes gradient corrections of order $\hbar^2$ in the particle
density and the scalar field. This new semiclassical approach has
already been applied to the study of finite nuclei
\cite{Cen93b,Cen92,Von92a,Cen93a} and of the nuclear surface in the
symmetric case \cite{Cen93b,Spe93,Cen93c}. In comparison with quantal
calculations, the inclusion of the gradient corrections to the TF
method improves the surface properties (energy and profile)
\cite{Cen92,Spe93,Cen93c}. Moreover, the quality of the RETF results
is less dependent on the parameters of the relativistic interaction
than in the TF approximation.

The asymmetric matter has also been investigated within the
relativistic theory. For example, the infinite system has been
analyzed in the Dirac--Hartree--Fock approach \cite{Lop88}, and
surface properties at low proton concentrations have been computed
with the TF method \cite{Von94b}. M\"uller and Dreizler \cite{Mul93}
have calculated relativistic hot asymmetric nuclear matter and nuclei,
and in Ref.\ \cite{Mul95} M\"uller and Serot have carefully dealt with
the subject of phase transitions in asymmetric nuclear matter at
finite temperature. Other works on relativistic asymmetric systems
have been concerned, e.g., with neutron star properties \cite{Eng94}
or collective modes and response functions \cite{Mat94}.

The main aim of this paper is to analyze, from a semiclassical point
of view, the surface and curvature properties of asymmetric
semi-infinite nuclear matter in the non-relativistic and relativistic
frameworks. For this purpose we will use the ETF and RETF methods
respectively. Our study covers the range of proton concentrations from
symmetric nuclear matter to beyond the proton drip. We will put
special emphasis on the calculation of the curvature energy, that can
be fully obtained only within the semiclassical approach and in which
some confusion exists \cite{Dur93,Cen96,Cen93c}.

The paper is organized as follows. Section 2 is devoted to the study
of uniform asymmetric nuclear matter with arbitrary proton
concentration, including drip neutrons and protons. In the third
section we derive the equations for the interface surface and
curvature properties in the general case with drip particles.
Numerical investigations are presented and discussed in Section 4. The
summary and conclusions are given in the last section.

\pagebreak
% >>>>>>>>>>>>>>>>>>>>>>>>>>>>>>>>>>>>>>>>>>>>>>>>>>>>>>>>>>>>>>>>>>>>
% THE BULK EQUILIBRIUM
%
\section{The bulk equilibrium}
\hspace*{\parindent}
Before proceeding to study the surface properties of nuclei in
coexistence with drip nucleons, we start by the simpler model of two
infinite pieces of asymmetric nuclear matter in equilibrium. The
uniform phase with the higher density and a given proton concentration
represents a nucleus that is in equilibrium with a surrounding gas of
drip nucleons. The latter is represented by the lower density phase,
with another proton concentration. This approach neglects the
interface effects and reflects the situation of the equilibrium
densities very far from the interface region in either direction.

The asymmetric nuclear matter problem including drip particles has a
notable similitude with the problem of the coexistence between nuclei
and evaporated nucleons at finite temperature
\cite{Mul93,Bar81,Bon85,Sur87}. Actually, the bulk equilibrium of
asymmetric matter at finite temperature was investigated time ago
within the astrophysical context \cite{Lat78,Bar80}, and more recently
in Refs.\ \cite{Mul95,Das92}. In this section we perform a similar
analysis at $T= 0$ based on the relativistic non-linear \sw model,
with care over the effects of the proton concentration. However, our
discussion is rather general and can be applied to the
non-relativistic treatment of cold asymmetric nuclear matter as well.

First we consider an infinite system made up of two components
(neutrons and protons) in a single phase. For describing asymmetric
matter it is convenient to switch variables from the neutron and
proton densities $\rho\dr{n}$ and $\rho\dr{p}$ to the total density
$\rho$ and the relative neutron excess $\delta$:
% Equation 1
\beq
\rho= \rho\dr{n} + \rho\dr{p}
\label{eq1} \eeq
%Equation 2
\beq
\delta= \frac{\rho\dr{n}-\rho\dr{p}}{\rho} \,.
\label{eq2}  \eeq
If we call $\ed(\rho,\delta)$ the energy density of the system, then
the neutron and proton chemical potentials and the pressure are
%
% Equation 3
\beqa
 \mu\dr{n}  =  \frac{\partial\ed(\rho,\delta)}{\partial\rho\dr{n}}
& = & \frac{\partial\ed}{\partial\rho}\frac{\partial
 \rho}{\partial\rho\dr{n}} + \frac{\partial\ed}{\partial\delta}
 \frac{\partial\delta}{\partial\rho\dr{n}}  =
 \frac{\partial\ed} {\partial\rho}
+ \frac{\partial\ed}{\partial\delta}\frac{1 - \delta} {\rho}
\label{eq3} \\[3mm]
%
% Equation 4
 \mu\dr{p}  =  \frac{\partial\ed(\rho,\delta)}{\partial\rho \dr{p}}
 & = & \frac{\partial\ed}{\partial\rho}\frac{\partial
 \rho}{\partial\rho\dr{p}} + \frac{\partial\ed}{\partial\delta}
 \frac{\partial\delta}{\partial\rho\dr{p}} =
  \frac{\partial\ed} {\partial\rho}
 - \frac{\partial\ed}{\partial\delta}\frac{1 +\delta} {\rho}
\label{eq4}\\[3mm]
%
% Equation 5
 P  = -\frac{\partial}{\partial\lf(1/{\rho}\ri)} \lf[\frac
{\ed(\rho,\delta)}{\rho}\ri] & = &
\rho^2\frac{\partial}{\partial\rho}\lf[\frac{\ed(\rho,\delta)}
{\rho}\ri] =
\rho\frac{\partial\ed}{\partial\rho} - \ed \,.
\label{eq5}
\eeqa
Manipulating these expressions, one gets that the pressure is related
with the neutron and proton chemical potentials through
%
% Equation 6
\beq
P = \mu\dr{n}\rho\dr{n} + \mu\dr{p}\rho\dr{p} - \ed(\rho,\delta) \,.
\label{eq6}\eeq
At saturation $P = 0$ and the preceding equation is, actually, a
generalization of the Hugenholtz--Van Hove theorem \cite{Sat83,Far86}.
The incompressibility at saturation of asymmetric nuclear matter with
relative neutron excess $\delta$ reads
%
%Equation 7
\beq
K(\rho\dr{sat},\delta) =
9\rho^2\frac{d^2}{d\rho^2}\lf.\lf(
\frac{\ed(\rho,
\delta)}{\rho}\ri)\ri|_{\rho\dr{sat}} ,
\label{eq7} \eeq
where $\rho\dr{sat}$ is the saturation density that conforms Eq.\
\req{eq6} with $P=0$. The expression of the energy density $\ed$ for
the relativistic model and for Skyrme forces can be found in Appendix
A\@. There, $\ed$ is given including the semiclassical gradient
corrections of order $\hbar^2$ which, of course, vanish in the uniform
infinite geometry.

The asymmetric nuclear matter in a single phase is not stable at all
densities and $\delta$ values. At zero temperature, the necessary and
sufficient conditions for the stability of a binary system can be
expressed through the inequalities \cite{Mul95,Bar80,Cal60}
%equation 8
\beq
K = \lf(\frac{\partial P}{\partial\rho}\ri)\dr{\delta} \geq 0
\label{eq8} \eeq
%
%equation 9
\beqa
\lf(\frac{\partial\mu\dr{n}}{\partial\delta}\ri)_{P} \geq 0
& \rm{or} &
 \lf(\frac{\partial\mu\dr{p}}{\partial\delta}\ri)_{P} \leq 0 \,.
\label{eq9} \eeqa
In addition, the pressure $P$ must be positive. A positive
incompressibility $K$ guarantees mechanical stability, and the
condition on the derivatives of the chemical potential ensures
diffusive stability. Violation of the stability criteria signals phase
separation. The system ceases to exist as a phase alone and splits
into separate phases in equilibrium. However, instability is only a
sufficient condition and phase separation may take place when the
single-phase system is stable (in general, metastable).

Now we address the problem of the coexistence of two phases in a
two-component system. One of the phases represents an asymmetric
uniform nuclear system with density $\rho_0$ and relative neutron
excess $\delta_0$. The other phase corresponds to a uniform phase
of drip particles characterized by $\rho\dr{d}$ and $\delta\dr{d}$.
Equation \req{eq6} is still valid for each separate phase, but at
equilibrium the pressure is no longer zero. In the general case with
drip neutrons and protons, the two phases can coexist when the
following conditions are fulfilled:
% equation 10
\beqa
 \mu\dr{n0} & = & \mu\dr{nd}
\label{eq10} \\[3mm]
%
% equation 11
 \mu\dr{p0} & = & \mu\dr{pd}
\label{eq11} \\[3mm]
%
% equation 12
 P_0 & = & P\dr{d}  \,,
\label{eq12}\eeqa
which in view of Eq.\ \req{eq6} result in
%
% Equation 12b
\beq
 \mu\dr{n} (\rho\dr{n0}-\rho\dr{nd}) +
 \mu\dr{p} (\rho\dr{p0}-\rho\dr{pd}) =
 \ed_0 - \ed\dr{d} \,.
\label{eq12b}\eeq
Equations \req{eq10}--\req{eq12} allow one to get $\rho_0$,
$\rho\dr{d}$ and $\delta\dr{d}$ for a fixed value of $\delta_0$ in
the nuclear phase. Notice that the solution of the above equations
does not correspond to the saturation conditions at $\delta_0$ and
$\delta\dr{d}$. As a consequence, the energy of the nuclear and drip
phases in coexistence is not a minimum.

The numerical results of this section will be discussed for the NL1
parametrization of the relativistic non-linear \sw model \cite{Rei86}.
Other \sw parameter sets or non-relativistic Skyrme forces
qualitatively show the same trends. The saturation properties of NL1
are given in Table 1, together with those of other interactions we
will utilize later in the calculation of surface properties.

To study the different regimes of phase coexistence, Fig.\ 1 displays
the neutron and proton chemical potentials against the relative
neutron excess for NL1. At small values of $\delta_0$ there are no
drip particles and the nuclear matter fulfills the usual saturation
condition. In this first stage the chemical potentials are negative.
While for neutrons the chemical potential is an increasing function of
$\delta_0$, it decreases steadily for protons. The point where the
neutron chemical potential vanishes marks the onset of the neutron
drip. At this moment the coexistence between the nuclear and drip
phases, the latter containing only neutrons ($\delta\dr{d}=1$), starts
to be possible. In this second regime, that for NL1 covers the range
$0.24 \leq \delta_0 \leq 0.64$ approximately, the neutron chemical
potential ($\mu\dr{n0}= \mu\dr{nd}$) is positive. Since there are no
protons in the drip matter yet, the proton chemical potentials
$\mu\dr{p0}$ and $\mu\dr{pd}$ of the nuclear and drip phases are
different. Both $\mu\dr{p0}$ and $\mu\dr{pd}$ are negative, being
lowered by raising the neutron excess. Due to the fact that
$\mu\dr{pd}$ is larger and decreases at a faster rate than
$\mu\dr{p0}$, they match at some large $\delta_0$, which for NL1
happens at $\delta_0 \simeq 0.64$. This is just the point where
protons start to drip. The last two rows in Table 1 display the values
of $\delta_0$ at the neutron and proton drip points (labelled
$\bar\delta\dr{nd}$ and $\bar\delta\dr{pd}$).

For larger $\delta_0$ the drip phase contains a small proton
concentration ($\delta\dr{d} < 1$). The two phases disappear when the
solution of Eqs.\ \req{eq10}--\req{eq12} yields the same density and
neutron excess in the nuclear and drip regions. This solution is just
the critical point characterized by $\rho\dr{c}$ and $\delta\dr{c}$,
see Eq.\ \req{eq13} below. It corresponds to the maximum of
$\mu\dr{n0}(\delta)$ and to the minimum of $\mu\dr{p0}(\delta)$. For
$\delta$ larger than $\delta\dr{c}$ the dashed lines in Fig.\ 1
represent the neutron and proton chemical potentials of the drip
matter only. Tracing horizontal lines in the $\mu-\delta$ plane one
can read, at the left and right hand sides of the critical value
$\delta\dr{c}$, the relative neutron excess of the nuclear and drip
phases in coexistence beyond the proton drip point.

Figure 2 shows the nuclear and drip densities, $\rho_0$ and
$\rho\dr{d}$, as a function of the square of the neutron excess for
NL1. For small values of $\delta_0$ the nuclear density $\rho_0$ has a
linear behaviour with $\delta_0^2$, as expected from the droplet model
\cite{Mye69}. The density of the nuclear (drip) phase falls (climbs)
with increasing $\delta_0^2$. Both densities coincide at the critical
$\delta\dr{c}$. The dashed line of Fig.\ 2 represents the density of
the drip region for $\delta^2 > \delta^2\dr{c}$. As in Fig.\ 1, by
drawing horizontal lines it serves to read the equilibrium values of
$\delta_0^2$ and $\delta\dr{d}^2$ when protons have migrated to the
drip phase.

The equation of state of NL1 at several values of $\delta$ is
displayed by solid lines in Fig.\ 3. The dot-dashed line defines the
boundary of the coexistence region. The long-dashed and short-dashed
lines are the diffusive and mechanical stability lines respectively.
We have joined by dashed horizontal lines a few examples of points on
the coexistence curve, belonging to the nuclear and drip phases, that
can be in equilibrium. The common maximum of the coexistence and
diffusive stability lines corresponds to the critical point
($\delta\dr{c}, \rho\dr{c}$) determined by
%%Equation 13
\beq
\lf(\frac{\partial\mu\dr{n}}{\partial\delta}\ri)_{P} =
\lf(\frac{\partial^2\mu\dr{n}}{\partial\delta^2}\ri)_{P} = 0
 \,.
\label{eq13} \eeq
The line of mechanical stability lies in the negative pressure region.
Therefore, it does not play any role in the separation of the system
in two phases, that is completely ruled by the diffusive instability.
The zone between the coexistence and diffusive stability lines is the
metastable region where the binary system may remain in a single phase
or undergo phase separation.

Consider a uniform nuclear system with density $\rho$ and neutron
excess $\delta$ under a pressure $P$. If the representative point in
the $P-\rho$ diagram of Fig.\ 3 lies outside the coexistence region,
the system is in a single phase. If the system is expanded along a
line of constant $\delta$ it will remain in a single phase up to
crossing the coexistence line, where phase separation can occur.
Denoting by $\xi$ the mass fraction of the nuclear phase ($\delta_0,
\rho_0$) in coexistence with the incipient drip phase ($\delta\dr{d},
\rho\dr{d}$), with mass fraction $1-\xi$, the following relationships
are to be satisfied at that point:
%
%Equation 14
\beqa
\delta & =& \xi\delta_0 + (1 - \xi)\delta\dr{d}
\label{eq14} \eeqa
%
%Equation 15
\beqa
\frac{1}{\rho} & = & \frac{\xi}{\rho_0} +
\frac{1-\xi}{\rho\dr{d}} \,.
\label{eq15} \eeqa

In Figs.\ 4 and 5 we plot in the $\mu\dr{n}-\rho$ and $\mu\dr{p}-\rho$
planes the lines of coexistence (solid), diffusive stability (long
dash) and mechanical stability (short dash) for NL1. The chemical
potentials and densities at the right and left hand sides of the
critical point (the maximum and the minimum of the coexistence line in
the $\mu\dr{n}-\rho$ and $\mu\dr{p}-\rho$ planes) correspond to the
nuclear and drip phases respectively. The dashed straight lines
joining points of the coexistence curve indicate some possible values
of the densities and chemical potentials in the phase equilibrium.
Notice that in the $\mu\dr{p}-\rho$ plane some of these straight lines
are not horizontal. In such cases the drip phase does not contain
protons ($\delta\dr{d}= 1$) and the proton chemical potentials of both
phases are different, see Fig.\ 1. When the proton drip point is
reached these lines turn horizontal, indicating that $\mu\dr{p0} =
\mu\dr{pd}$.

\pagebreak
% >>>>>>>>>>>>>>>>>>>>>>>>>>>>>>>>>>>>>>>>>>>>>>>>>>>>>>>>>>>>>>>>>>>>
% SURFACE PROPERTIES IN TWO-COMPONENT SYSTEMS.
%
\section{Surface properties in two-component systems}
\hspace*{\parindent}
On the basis of asymmetric semi-infinite nuclear matter, in this
section we shall study the surface effects of a neutron-rich nucleus
immersed in a gas of drip particles. This regime corresponds to the
physical situation found in neutron stars, at densities between $4
\times 10^{11}$ g/cm$^3$ and normal nuclear values ($2.7 \times
10^{14}$ g/cm$^3$) \cite{Rav72}. First of all it is necessary to
determine the density profile from which the surface and curvature
energies can be obtained. To do that one considers a semi-infinite
slab with a plane interface separating a mixture of protons and
neutrons to the left and a gas of drip particles to the right. The
axis perpendicular to the interface is taken to be the $z$ axis. The
density profile is sketched in Fig.\ 6. Observe that the relative
neutron excess $\delta$ depends on the $z$ coordinate. When $z$ goes
to minus infinity, the neutron and proton densities approach the
values that in uniform nuclear matter are in equilibrium with a
uniform gas of drip particles. Therefore, for a given bulk neutron
excess $\delta_{0} \equiv \delta(-\infty)$, the values of
$\rho(-\infty)$, $\rho(\infty)$ and $\delta(\infty)$ are just the
$\rho_0$, $\rho\dr{d}$ and $\delta\dr{d}$ solutions to Eqs.\
\req{eq10}--\req{eq12}.

The problem of finding the density profile at a certain value of
$\delta_{0}$ when drip particles appear is, in a sense, formally
equivalent to solving the problem of a warm nucleus. In that case, as
it was pointed out by Bonche {\it et al.}\ \cite{Bon85}, the
Hartree--Fock equations at some temperature and chemical potential
possess two different solutions. One of them represents the nuclear
system in equilibrium with its evaporated nucleons. The other one
belongs to the evaporated nucleon gas alone. It is then natural to
define extensive magnitudes characterizing the nucleus as the {\em
difference} between their value in the nuclear-plus-gas solution and
their value in the gas solution. The same idea has been applied in the
semiclassical scheme for the Euler--Lagrange equations at finite
temperature \cite{Sur87}.

To compute semiclassically the proton and neutron densities in
asymmetric semi-infinite nuclear matter one has to minimize the total
energy per unit area. This has to be done with the constraint of
conservation of the number of neutrons and protons with respect to
arbitrary variations of the densities. According to the above
discussion, the energy, neutron and proton densities entering the
constrained energy of the (semi-infinite) nucleus will be taken as the
difference between the ones of the whole system (nucleus plus drip
particles), and those of the uniform gas of drip particles (indicated
by a {d} subscript). For instance we will write the local energy
density representing the nucleus, i.e.\ the subtracted system, in the
form $\ed(z) - \ed\dr{d}$. As $z \rightarrow -\infty$ we have $\ed(z)
\rightarrow \ed_0$, while as $z \rightarrow \infty$ it is $\ed(z)
\rightarrow \ed\dr{d}$. Due to the absence of Coulomb forces in the
semi-infinite problem, the drip particles are described by plane waves
and behave as a dilute uniform phase.

Then, we write the constrained energy per unit area as
% equation 16
\beq
 \frac{E\dr{const}}{S} = \int_{-\infty}^\infty dz \lf\{\ed (z)-
\ed\dr{d} - \mu\dr{n}[\rho\dr{n}(z) - \rho\dr{nd}] -
\mu\dr{p}[\rho\dr{p}(z) - \rho\dr{pd}]\ri\} .
\label{eq16} \eeq
For the whole system, the proton and neutron densities are obtained
from the following coupled Euler--Lagrange equations:
%Equation 17,18
\beq
\frac{\delta \ed (z)}{\delta \rho\dr{n}}- \mu\dr{n} = 0 \,,
\hspace{1cm}
\frac{\delta \ed (z)}{\delta \rho\dr{p}}- \mu\dr{p} = 0 \,.
\label{eq18}
\eeq
In the relativistic framework one has three additional equations,
originating from the variations with respect to the fields $\phi$, $V$
and $R$ associated with the $\sigma$, $\omega$ and $\rho$ mesons:
%
%Equation 19,20,21
\beq
\frac{\delta \ed (z)}{\delta \phi }= 0 \,,
\hspace{1cm}
\frac{\delta \ed (z)}{\delta V }  = 0 \,,
\hspace{1cm}
\frac{\delta \ed (z)}{\delta R }  = 0 \,.
\label{eq20}
\eeq
We recall the reader that the ETF and RETF expressions for $\ed$ are
detailed in Appendix A\@. The drip particles obey analogous equations:
%
%Equation 21,22
\beq
\frac{\delta \ed\dr{d}}{\delta \rho\dr{nd}}- \mu\dr{n}= 0 \,,
\hspace{1cm}
\frac{\delta \ed\dr{d}}{\delta \rho\dr{pd}}- \mu\dr{p} = 0 \,,
\label{eq22}
\eeq
plus in the relativistic case
%Equation 23,24
\beq
\frac{\delta \ed\dr{d}}{\delta \phi\dr{d}}= 0 \,,
\hspace{1cm}
\frac{\delta \ed\dr{d}}{\delta V\dr{d}}  = 0 \,,
\hspace{1cm}
\frac{\delta \ed\dr{d}}{\delta R\dr{d}}  = 0 \,.
\label{eq24}
\eeq

The local equations \req{eq18} and \req{eq20} are self-consistently
solved by numerical iteration. As explained in Ref.\ \cite{Cen93b}, we
employ the imaginary time-step method to get the densities from Eq.\
\req{eq18} and Gaussian elimination to get the meson fields from Eq.\
\req{eq20}. Given $\delta_0$, the asymptotic boundary conditions to be
imposed stem from the coexistence equations \req{eq10}--\req{eq12}.
Equations \req{eq22} and \req{eq24} are satisfied by the low-density
phase solution of \req{eq10}--\req{eq12}, which provides the limiting
behaviour as $z \rightarrow \infty$. On the other hand, the
high-density solution of \req{eq10}--\req{eq12} represents the system
at $z= -\infty$.

Once the density profiles are known, the next step is to get the
surface and curvature energy coefficients. The surface tension (i.e.\
the surface energy per unit area of a flat surface) may be written as
\cite{Mye69,Mye85,Kol85}
%Equation 25
\beq
\sigma= \int_{-\infty}^{\infty} dz
\lf[\ed (z)-\ed\dr{d} - \ed\dr{ref}(z)\ri] ,
\label{eq25}
\eeq
where $\ed\dr{ref}(z)$ is a reference energy density whose integral
represents in some way the bulk contribution. Myers {\it et al.}\
\cite{Mye85} noted that there are two possibilities to define
$\ed\dr{ref}(z)$.

The first definition corresponds to a reference energy that represents
the energy a nucleus would have if its nucleons would belong to
infinite nuclear matter. This reference energy density is to be
written as
%
% equation 26
\beq
\ed\ur{e}\dr{ref}(z)=
\frac{\ed_0 -\ed\dr{d}}
{\rho_0 - \rho\dr{d}}\lf[\rho (z) - \rho\dr{d} \ri] ,
\label{eq26}\eeq
which we will call the e-definition following Ref.\ \cite{Mye85}. In
this case the surface tension reads
%
%Equation 27
\beq
\sigma\dr{e} = \int_{- \infty}^\infty dz \lf\{ \ed (z) - \ed\dr{d} -
\frac{\ed_0 - \ed\dr{d}}{\rho_0
-\rho\dr{d}}\lf[\rho (z) - \rho\dr{d}\ri]\ri\} .
\label{eq27}\eeq

The second definition of the reference energy will be called the Gibbs
definition. Instead of the bulk energy per particle, it introduces the
neutron and proton chemical potentials and the pressure associated
with the bulk. The Gibbs reference energy for the whole system is
% equation 28
\beq
E^\mu\dr{ref} = \mu\dr{n}N + \mu\dr{p}Z - P V ,
\label{eq28} \eeq
while for the drip phase we have
% equation 29
\beq
E^\mu\dr{ref,d} =
 \mu\dr{nd}N\dr{d} + \mu\dr{pd}Z\dr{d} - P\dr{d}V .
\label{eq29}\eeq
The meaning of $E^\mu$ becomes evident if one writes the
thermodynamic relation
%
%Equation 30
\beq
dE= \mu\dr{n}dN + \mu\dr{p}dZ - PdV \,,
\label{eq30} \eeq
where $dE$ is the energy necessary to remove $dN$ neutrons, $dZ$
protons and reduce the volume by an amount $dV$. Thus, $E^\mu$ may
be interpreted as a reference disassembly energy \cite{Mye85}.

With the help of the equilibrium conditions \req{eq10}--\req{eq12}
between the nuclear and drip phases, the reference energy for the
subtracted system reads
%Equation 31
\beqa
E^\mu\dr{ref} - E^\mu\dr{ref,d} & = & \mu\dr{n}(N -
N\dr{d}) + \mu\dr{p}(Z -Z\dr{d})
\nonumber \\[3mm]
& = &
\int_{-\infty}^\infty dz \lf\{ \mu\dr{n}\lf[\rho\dr{n}(z)
-\rho\dr{nd}\ri] + \mu\dr{p}\lf[\rho\dr{p}(z) -\rho\dr{pd}\ri]\ri\} .
\label{eq31} \eeqa
From this equation we can identify a Gibbs reference energy density:
%
%Equation 31bis
\beq
 \ed\dr{ref}^\mu(z) =
 \mu\dr{n}\lf[ \rho\dr{n}(z) - \rho\dr{nd}\ri] +
 \mu\dr{p}\lf[ \rho\dr{p}(z) -\rho\dr{pd}\ri] .
\label{eq31b} \eeq
Substituting Eq.\ \req{eq31b} into Eq.\ \req{eq25}, the Gibbs surface
tension becomes
%
%Equation 32
\beq
\sigma_\mu=\int_{-\infty}^\infty dz \lf\{\ed (z) - \ed\dr{d} -
\mu\dr{n}[\rho\dr{n}(z)-\rho\dr{nd}] -
\mu\dr{p}[\rho\dr{p}(z)-\rho\dr{pd}] \ri\} .
\label{eq32} \eeq
As a side remark we note that $\ed\dr{ref}^\mu (-\infty) =
\ed\dr{ref}\ur{e} (-\infty) = \ed_0 -\ed\dr{d}$ and that
$\ed\dr{ref}^\mu (\infty) = \ed\dr{ref}\ur{e} (\infty) = 0$.

Comparing with Eq.\ \req{eq16} one concludes that the surface tension
that must be minimized for obtaining the density profile is the Gibbs
surface tension \req{eq32}. Attempts to minimize the surface tension
$\sigma\dr{e}$, Eq.\ \req{eq27}, result in the non-conservation of
neutrons and protons \cite{Far86}. Nevertheless, $\sigma\dr{e}$ can be
calculated from the variational densities arising from the
minimization of $\sigma_\mu$. Of course, in the case of symmetric
semi-infinite nuclear matter, $\sigma\dr{e}$ and $\sigma_\mu$ coincide
owing to the Hugenholtz--Van Hove theorem. The surface energy
coefficients associated with $\sigma\dr{e}$ and $\sigma_\mu$ are given
by
%
%Equation 33
\beqa
E\dr{s,e}   & = & 4\pi r_0^2\sigma\dr{e}
\label{eq33}
\\
%Equation 34
E\dr{s,\mu} & = & 4\pi r_0^2\sigma_\mu \,,
\label{eq34} \eeqa
where $r_0$ is the nuclear radius constant:
%
%Equation 35
\beq
\frac{4}{3}\pi r_0^3\lf(\rho_0 -\rho\dr{d}\ri) = 1 \,.
\label{eq35} \eeq

Another useful quantity for our study is the so-called surface
location $z_0$. It is defined as \cite{Mye85}
%
% Equation 36
\beq
z_0 = \frac{\int_{-\infty}^\infty z\rho^\prime(z) dz}
{\int_{-\infty}^\infty \rho^\prime(z) dz} \,,
\label{eq36}\eeq
where the prime denotes a derivative with respect to $z$. Similar
expressions hold for the neutron and proton surface locations
($z\dr{0n}$ and $z\dr{0p}$). It is easy to show that
%
% Equation 37
\beq
 z_0 \Delta\rho =
 z\dr{0n} \Delta\rho\dr{n} + z\dr{0p} \Delta\rho\dr{p} \,,
\label{eq37}\eeq
with $\Delta\rho \equiv \rho_0 - \rho\dr{d}$, $\Delta\rho\dr{n} \equiv
\rho\dr{n0} - \rho\dr{nd}$ and $\Delta\rho\dr{p} \equiv \rho\dr{p0} -
\rho\dr{pd}$.
After some algebra one obtains an expression relating the two
definitions of the surface tension:
% Equation 38
\beq
\sigma\dr{e}-\sigma_\mu = (\mu\dr{n} - \mu\dr{p}) (z\dr{0n} -
z\dr{0p})\frac{\Delta\rho\dr{n}  \Delta\rho\dr{p}}{\Delta\rho} \,.
\label{eq38}\eeq

In semi-infinite nuclear matter the curvature energy coefficient
$E\dr{c}$ is given by \cite{Mye69,Sto73,Sto85}
% Equation 39
\beqa
E\dr{c} & = & E\dr{c}\ur{geo} + E\dr{c}\ur{dyn}
\nonumber \\[3mm]
E\dr{c}\ur{geo} & = &
 8\pi r_0 \int_{-\infty}^\infty dz (z - z_0)
\lf[ \ed(z) - \ed\dr{d} -\ed\dr{ref}(z)\ri]
\nonumber \\[3mm]
E\dr{c}\ur{dyn} & = &
 8\pi r_0 \int_{-\infty}^\infty dz
\lf. \frac{\partial}{\partial \kappa}
\lf[\ed(z) - \ed\dr{d} - \ed\dr{ref}(z)\ri] \ri|\dr{\kappa=0} ,
\label{eq39}\eeqa
where $\kappa$ is the curvature ($\kappa= 2/R$ for a sphere of radius
$R$). The two contributions to the curvature energy coefficient in
Eq.\ \req{eq39} are called geometrical ($E\dr{c}\ur{geo}$) and
dynamical ($E\dr{c}\ur{dyn}$) respectively. The geometrical
contribution only involves the variation of the surface energy density
$\ed(z) - \ed\dr{d} -\ed\dr{ref}(z)$ across the surface parallel to
the $z$ axis. The dynamical part comes from the change of the surface
energy density by curvature when the plane surface is infinitesimally
bent.

In order to establish a connection between $E\ur{geo}\dr{c,e}$ and
$E\ur{geo}\dr{c,\mu}$ it is helpful to define the surface width of the
density:
%
% Equation 40
\beq
b^2 = \frac {\int_{-\infty}^\infty (z - z_0)^2
\rho^\prime(z) dz}{\int_{-\infty}^\infty
\rho^\prime(z) dz} \,,
\label{eq40}\eeq
plus similar quantities $b\dr{n}$ and $b\dr{p}$ for the neutron and
proton densities. One finds that $b$, $b\dr{n}$ and $b\dr{p}$ satisfy
%
% Equation 41
\beq
b^2 \lf(\Delta\rho\ri)^2 =
  b\dr{n}^2 \, \Delta\rho\dr{n} \Delta\rho
+ b\dr{p}^2 \, \Delta\rho\dr{p} \Delta\rho
+ (z\dr{0n}-z\dr{0p})^2 \, \Delta\rho\dr{n} \Delta\rho\dr{p} \,.
\label{eq41} \eeq
Making some calculations it can be shown that
%
% Equation 42
\beq
\frac{E\dr{c,e}\ur{geo} - E\dr{c,\mu}\ur{geo}}{8\pi r_0} =
\frac{(\mu\dr{n}-\mu\dr{p})\Delta\rho\dr{n}\Delta\rho\dr{p}}
{2\Delta\rho}\lf[b\dr{n}^2-b\dr{p}^2 - (z\dr{0n} - z\dr{0p})^2
\, \frac{\Delta\rho\dr{n}-\Delta\rho\dr{p}}{\Delta\rho}\ri] .
\label{eq42}\eeq

The dynamical part of the curvature energy requires paying a special
attention. In general, the surface energy density $\ed(z) - \ed\dr{d}
-\ed\dr{ref}(z)$ depends on the curvature $\kappa$ in two different
manners. First, $\ed(z)$ may have an explicit dependence on the
laplacian operator $\Delta$ which in the limit $R\rightarrow \infty$
reads as $d^2/dz^2 + \kappa d/dz$. Second, the particle
densities (and the meson fields in the relativistic case) carry an
implicit dependence on the curvature $\kappa$. By construction the
energy density entering Eq.\ \req{eq39} is free from any explicit
dependence on the Laplace operator (see Appendix A). Actually, this
dependence has been removed by partial integrations in the
semiclassical functionals \cite{Cen93b,Cen93c}. Consequently, in the
present calculation, the sole contribution we may have to the
dynamical curvature energy comes from the implicit curvature
dependence of the nuclear densities and fields. For non-relativistic
interactions we find
%
%Equation 44
\beqa
 E\dr{c}\ur{dyn} & = & 8\pi r_0 \sum_{q = {\rm n,p}}
\int^\infty_{-\infty} dz \lf\{\frac{\delta}{\delta \rho_q}
\lf[\ed(z) - \ed\dr{d} - \ed\dr{ref}(z) \ri] \frac{d\rho_q}{d\kappa}
\ri. \nonumber \\
& & \mbox{}
+ \lf. \lf. \frac{\delta}{\delta \rho_{q {\rm d}}}
\lf[\ed(z) - \ed\dr{d} - \ed\dr{ref}(z)\ri]
\frac{d\rho_{q {\rm d}}}{d\kappa} \ri\} \ri|_{\kappa=0} .
\label{eq44} \eeqa
In the relativistic case one has a similar expression but including
the variations with respect to the meson fields \cite{Cen96,Cen93c}.

If we insert in Eq.\ \req{eq44} the Gibbs reference energy
$\ed\dr{ref}^\mu (z)$, Eq.\ \req{eq31b}, the prefactors of the
derivatives with respect to $\kappa$ vanish by virtue of the
variational equations obeyed by the self-consistent densities (and
meson fields), cf.\ \req{eq18}--\req{eq24}. Therefore, in our
calculation the Gibbs dynamical curvature energy is zero
($E\dr{c,\mu}\ur{dyn}= 0$) and the full Gibbs curvature energy
coincides with its geometrical part: $E\dr{c,\mu}=
E\dr{c,\mu}\ur{geo}$.

This is not the situation when one takes the e-definition of the
reference energy $\ed\dr{ref}\ur{e}(z)$, Eq.\ \req{eq26}. In this case
one finds a non-zero dynamical curvature energy $E\dr{c,e}\ur{dyn}$,
because the prefactors of the derivatives on $\kappa$ in Eq.\
\req{eq44} do not vanish. ($E\dr{c,e}\ur{dyn}= 0$ only for symmetric
matter with $\delta_0= 0$.) Notice that in the relativistic framework
there are no contributions from the meson fields to
$E\dr{c,e}\ur{dyn}$ (as long as we employ the functionals of Appendix
A), since they do not enter in the reference energy.

$E\dr{c,e}\ur{dyn}$ can be recast in the form
%Equation 45
\beqa
E\dr{c,e}\ur{dyn} & = & 8\pi r_0 \int^\infty_{-\infty} dz
\lf. \frac{\partial}{\partial \kappa}\lf[\ed (z) - \ed\dr{d}
 - \ed\dr{ref}^\mu (z) \ri]\ri|_{\kappa= 0}
\nonumber \\[3mm]
& & \mbox{}
+ 8\pi r_0 \int^\infty_{-\infty}dz \lf. \frac{\partial}
{\partial \kappa}\lf[\ed\dr{ref}^\mu(z) -\ed\dr{ref}\ur{e}(z) \ri]
 \ri|_{\kappa=0} .
\label{eq45} \eeqa
The first term is just $E\dr{c,\mu}\ur{dyn}$, that vanishes according
to the foregoing discussion. As a consequence, the dynamical part of
the e-curvature energy can be expressed as
%Equation 46
\beq
 E\dr{c,e}\ur{dyn} = 8\pi r_0  \sum_{q = {\rm n,p}}
\lf[\mu_q - \frac{\ed_0 -\ed\dr{d}}{\rho_0 -\rho\dr{d}} \ri]
\int^\infty_{-\infty} dz  \lf. \frac{\partial}{\partial \kappa}
 \lf[\rho_q (z) - \rho_{q {\rm d}}\ri] \ri|_{\kappa=0} .
\label{eq46} \eeq

In general the evaluation of $\partial \rho_q / \partial \kappa$ in
Eq.\ \req{eq46} is not a trivial matter, e.g.\ it cannot be calculated
in a fully quantal way \cite{Cen96}. Fortunately, this problem can be
solved within the semiclassical formalism \cite{Dur93,Cen96}. The
reason is that the semiclassical expression of the particle density
derived from the Wigner--Kirkwood $\hbar$ expansion of the density
matrix \cite{Rin80} contains Laplace operators, and thus a dependence
on the curvature $\kappa$. From the Wigner--Kirkwood density,
following Refs.\ \cite{Dur93} and \cite{Cen96}, the dynamical
curvature energy $E\dr{c,e}\ur{dyn}$ can be calculated in the ETF and
RETF approaches. This is summarized in Appendix B\@.

\pagebreak
% >>>>>>>>>>>>>>>>>>>>>>>>>>>>>>>>>>>>>>>>>>>>>>>>>>>>>>>>>>>>>>>>>>>>
% DISCUSSION OF RESULTS.
%
\section{Discussion of Results}
\hspace*{\parindent}
To illustrate our results, we have chosen the Skyrme forces SkM*
\cite{Bar82} and SIII \cite{Bei75} as representative interactions for
the non-relativistic case, and the non-linear parameter sets NL1
\cite{Rei86}, NL2 \cite{Lee86} and NL-SH \cite{Sha93} in the
relativistic model. Table 1 collects the saturation properties of
nuclear matter for these interactions, plus the droplet model
parameters $J$ and $L$:
%
%Equation 51
\beqa
 J & = & \frac{1}{2 \rho}
 \lf. \frac{\partial^2 \ed(\rho,\delta) }{\partial \delta^2}
 \ri|_{\delta= 0, \rho= \rho\dr{nm}}
\nonumber \\[3mm]
 L & = & \frac{3 \rho}{2} \,
 \frac{\partial}{\partial \rho} \frac{\partial^2}{\partial \delta^2}
 \! \lf. \lf( \frac{\ed(\rho,\delta)}{\rho}\ri)
 \ri|_{\delta= 0, \rho= \rho\dr{nm}} ,
\label{eq51} \eeqa
with $\rho\dr{nm}$ the saturation density of symmetric nuclear matter.
Also given in Table 1 are the values $\bar\delta\dr{nd}$ and
$\bar\delta\dr{pd}$ of the bulk neutron excess at the neutron and
proton drip points.

Although SkM* and SIII resemble in the energy per particle, effective
mass $m^*_\infty/m$ ($\simeq 0.8$) and bulk symmetry energy $J$
($\simeq 30$ MeV), these forces differ mainly in the incompressibility
modulus $K$ (217 MeV for SkM* and 355 MeV for SIII). In the
relativistic case, NL1, NL2 and NL-SH also disagree in the value of
$K$: NL1 is similar to SkM*, NL-SH to SIII, and NL2 has the largest
$K$ ($\simeq 400$ MeV). The NL1 and NL2 sets have a higher bulk
symmetry energy $J$ ($\simeq$ 45 MeV) than the Skyrme forces, whereas
in NL-SH the value of $J$ is relatively close to SkM*. Concerning the
effective mass, in the relativistic sets it is small as compared with
the Skyrme forces. For NL1 and NL-SH $m^*_\infty/m$ is similar
($\simeq 0.6$), while for NL2 it is slightly larger. In any case,
notice that the effective mass has a different origin in the
relativistic than in the non-relativistic model \cite{Jam89}. All the
forces are able to give a reasonably good description of finite nuclei
in spite of their differences. In particular, the relativistic set
NL-SH is very well suited for nuclei near the neutron drip line
\cite{Sha93}, that cannot be described so well with other \sw sets or
Skyrme forces.

%
% >>>>>>>>>>>>>>>>>>>>>>>>>>>>>>>>>>>>>>>>>>>>>>>>>>>>>>>>>>>>>>>>>>>>
% The surface properties.
%
\subsection{The surface properties}
\hspace*{\parindent}
Figure 7 displays the neutron and proton local density profiles
obtained from the solution to the Euler--Lagrange equations
\req{eq18}--\req{eq24} for NL1 and SkM*. They are plotted for several
values of the relative neutron excess in nuclear matter $\delta_0$.
The separation between vertical bars is the surface thickness $t$,
defined as the 90\% to 10\% fall-off distance. At the top of the
figure we show the result for $\delta_0= 0$, the symmetric
semi-infinite geometry. The neutron and proton profiles coincide and
the surface region is roughly centered around $z = 0$. The second part
of the figure corresponds to $\delta_0 = 0.2$ for which the
density profiles begin to differ. The surface thickness for neutrons
and protons is very similar to the $\delta_0 = 0$ case. However,
while $t$ is still centered around $z = 0$ for the neutron profile,
for the proton density $t$ is shifted to negative values of $z$ owing
to the symmetry terms of the interaction. The third part of the figure
corresponds to $\delta_0= 0.4$, when drip neutrons have appeared both
in NL1 and in SkM*. The surface thickness of the neutron and proton
distributions has clearly increased. The density of the drip phase is
considerably higher for NL1 than for SkM*, as one would expect from
the fact that the neutron drip starts earlier for NL1 (Table 1).
Finally, at the bottom of Fig.\ 7 the profiles are plotted for a high
$\delta_0$ beyond the proton drip point. The surface thickness becomes
quite large because the interior density and the density of the drip
phase are very close.

Figures 8 and 9 offer a more detailed analysis of the surface. They
display the surface thickness $t$ and the surface width $b$, Eq.\
\req{eq41}, for neutrons and protons calculated with the relativistic
sets NL1 and NL2 and with the Skyrme forces SkM* and SIII\@. One can
see that the slope of the $t$ and $b$ curves as a function of
$\delta_0$ is steeper for neutrons than for protons. After the neutron
drip point, one observes the appearance of a relative maximum (whose
height strongly depends on the force) in the surface thickness and
width of the neutron density distribution. The presence of this
maximum has already been reported in earlier literature \cite{Mye85}.
From a qualitative point of view it can be understood as follows. If
the bulk symmetry coefficient $J$ is large (as in the \sw sets NL1 and
NL2), it costs the system a great amount of energy to produce an
asymmetry in the bulk, thus favouring the ejection of neutrons to the
surface region. However, for a given $\delta_0$ the neutron density in
the drip region is fixed by the coexistence equations
\req{eq10}--\req{eq12}. Therefore, if the neutrons pushed out by the
symmetry term of the force cannot accommodate in the drip region, they
concentrate at the surface and contribute to the development of a
maximum in $t$ and $b$. This fact is illustrated in Fig.\ 10, where
the number of neutrons per unit area in the surface region $N/S$ is
drawn as function of $\delta_0$ for NL2. We have calculated $N/S$ as
the integral of the neutron density $\rho\dr{n}(z) - \rho\dr{nd}$
within the 90\%--10\% fall-off distance. One recognizes that it is
immediately after the neutron drip point ($\bar\delta\dr{nd}= 0.222$
for NL2) when the greatest accumulation of neutrons in the surface
takes place.

Another quantity of interest for inspecting the surface of the
asymmetric system is the neutron skin thickness:
%
%Equation 47
\beq
\Theta = z\dr{0n} - z\dr{0p} \,,
\label{eq47} \eeq
where $z\dr{0n}$ and $z\dr{0p}$ are the neutron and proton surface
locations defined through Eq.\ \req{eq36}. Figure 11 depicts the
change of $\Theta$ against $\delta_0$ for our representative forces.
For small values of $\delta_0$ the neutron skin thickness behaves
linearly, as predicted by the liquid droplet model (LDM) \cite{Mye69}:
%
%Equation 48
\beq
\Theta = \frac{3 r_0}{2} \frac{J}{Q} \delta_0 \,.
     \label{eq48} \eeq
The value of $Q$, which measures the stiffness of the system against
pulling $z\dr{0n}$ and $z\dr{0p}$ apart, can be obtained from the
slope of $\Theta$ at $\delta_0 = 0$. At small $\delta_0$,
forces having a large $J/Q$ ratio have also a large neutron skin
thickness. As seen from Fig.\ 11, for larger $\delta_0$ the
neutron skin thickness starts to depart from the linear behaviour and
develops a maximum in the region beyond the neutron drip point. This
maximum, which also appears in former analyses with non-relativistic
forces \cite{Mye85,Kol85}, is more pronounced for the relativistic
interactions we study here. It is related with the accumulation of
neutrons at the surface we have discussed before. With further
increase in $\delta_0$ the neutron skin decreases, since the inside
and the outside matter become more and more alike and the surface is
washed out. At the critical $\delta\dr{c}$ one expects $\Theta$ to
eventually vanish.

Figure 12 displays $\sigma\dr{e}$ and $\sigma_\mu$ against
$\delta_0^2$. The surface tension $\sigma\dr{e}$ grows with
$\delta_0^2$ and reaches a maximum at some point beyond the
neutron drip. Then it falls off to zero as the limiting situation of a
uniform system is approached. The Gibbs surface tension shows a
different behaviour and it decreases monotonically from its value in
the symmetric case to zero. In Figs.\ 13 and 14 we present our results
for the $\gamma\dr{e}$ and $\gamma_\mu$ curvature energies per unit
length ($\gamma= E\dr{c}/8\pi r_0$). These quantities exhibit a
similar behaviour as a function of $\delta_0^2$ to the surface
tensions. NL1 presents a visible peak in $\gamma\dr{e}$. The dynamical
contribution $\gamma\dr{e}\ur{dyn}= E\dr{c,e}\ur{dyn}/8\pi r_0$ is
shown by dashed lines in Figs.\ 13 and 14. Remember that in our
calculation the dynamical curvature energy only differs from zero in
the e-definition for $\delta_0 \neq 0$. At high values of $\delta_0^2$
most of the e-curvature energy comes from the dynamical part.

For small values of $\delta_0$ the nuclear droplet model predicts
%
%Equation 49
\beq
E\dr{s,e} = E\dr{s,0} + \lf(\frac{9J^2}{4Q} +
\frac{2E\dr{s,0}L}{K} \ri) \delta_{0}^2
     \label{eq49} \eeq
%Equation 50
\beq
 E\dr{s,\mu} = E\dr{s,0} - \lf(\frac{9J^2}{4Q}
-\frac{2E\dr{s,0}L}{K} \ri) \delta_0^2  \,.
     \label{eq50} \eeq
$E\dr{s,0}$ is the surface energy coefficient for the symmetric case,
$K$ is the bulk incompressibility and $L$ reads for the LDM
coefficient defined in Eq.\ \req{eq51} that gives the density
dependence of the symmetry energy. As we realize from Eqs.\ \req{eq49}
and \req{eq50}, the symmetry contribution to the surface energy in the
LDM is positive in the case of the e-definition and negative with the
Gibbs prescription. This contribution consists of two terms. The main
one ($9J^2 /4Q$), represents the variation of the bulk symmetry energy
$J$ when the nucleus increases its neutron skin against the resistance
provided by the surface stiffness $Q$. The corrective term
$2E\dr{s,0}L/K$ describes the change of the volume energy produced by
a change in the bulk density \cite{Mye69}. While both terms
participate additively in the symmetry contribution to $E\dr{s,e}$,
they have opposite signs in $E\dr{s,\mu}$. Since $\sigma_e$ and
$\sigma_{\mu}$ must behave according to Eqs.\ \req{eq49} and
\req{eq50} for small $\delta_0^2$, and they must vanish for the
uniform system, one can qualitatively understand the global trends of
Fig.\ 12. Apart from using Eq.\ \req{eq48} to obtain the
surface-stiffness coefficient $Q$, it can be extracted from the slope
of the difference $E\dr{s,e} - E\dr{s,\mu}$ that for small values of
$\delta_0$ behaves as
%
%Equation 52
\beq
 E\dr{s,e} - E\dr{s,\mu} = \frac{9J^2}{2Q} \delta_0^2 \,.
\label{eq52} \eeq

Table 2 collects $E\dr{s}$, $E\dr{c}$ and $t$ computed in the
symmetric case ($\delta_0 = 0$) and the coefficient $Q$, for SkM*,
SIII, NL1, NL2 and NL-SH\@. For the sake of comparison, Table 2 shows
in addition results from Refs.\ \cite{Kol85} and \cite{Von94b}. Our
values of the surface-stiffness coefficient calculated from Eqs.\
\req{eq48} and \req{eq52} are in good agreement between them, though
small differences arise. In general, the value of $Q$ from Eq.\
\req{eq48} is $\sim 0.5-1$ MeV greater than the value from Eq.\
\req{eq52}.

The strength of the peaks that appear in $t\dr{n}$, $b\dr{n}$ and
$\Theta$ in Figs.\ 8, 9 and 11 is, actually, related with the value of
the surface-stiffness coefficient of the force. Comparing with Table
2, one can see that forces with small values of $Q$ have the peaks
more developed. As discussed above, such peaks are connected with the
neutrons pushed to the surface. Since $Q$ measures the resistance
against separating neutrons from protons, it is qualitatively clear
that forces having a small $Q$ will tend to concentrate more neutrons
at the surface. A similar dependence on $Q$ is observed in the height
of the maxima presented by $\sigma\dr{e}$ and $\gamma\dr{e}$ in Figs.\
12--14.

From Table 2 we can see that non-relativistic and relativistic
parametrizations are able to give comparable surface properties.
Nevertheless, for the symmetric case, in Ref.\ \cite{Cen93c} a
tendency to thinner density profiles was noticed in the relativistic
model, with smaller values of the thickness $t$ when the relativistic
interaction was adjusted to give the surface energy of Skyrme forces.
Figure 8 suggests that this trend may be reversed at large
asymmetries, where $t$ (and $b$ as well, Fig.\ 9) grows faster with
$\delta_0$ for NL1 and NL2 than for SkM* and SIII\@. To get more
insight, we have performed calculations with the relativistic set
named RSk1* in Ref.\ \cite{Cen93c}. The parameters of RSk1* were
fitted to the nuclear matter properties of SkM* (including the
coefficient $J$), and the scalar mass was chosen to obtain the surface
energy of SkM* at $\delta_0= 0$ in the ETF approach. For RSk1* we find
$L= 81.8$ MeV and $Q \approx 25$ MeV, whereas for SkM* it is $L= 45.8$
MeV and $Q \approx 39$ MeV\@. Observe from this and Tables 1 and 2 an
overall tendency of the relativistic sets to have larger $L$ and
smaller $Q$ than the Skyrme forces.

Figure 15 summarizes our results for the surface thickness and surface
tension with SkM* and RSk1*. The relativistic thickness $t$ is smaller
for low neutron excess. But it increases more rapidly with $\delta_0$
and becomes clearly greater than for SkM* at high asymmetries, meaning
that the interface disappears sooner in the relativistic case. The
height of the relative maximum developed by the surface thickness of
the neutron density is larger for RSk1* ($Q \approx 25$ MeV) than for
SkM* ($Q \approx 39$ MeV). In Fig.\ 15 the curves of the Gibbs surface
tension $\sigma_\mu$ for RSk1* and SkM* closely follow each other
($\sigma_\mu$ is the minimized quantity in the semi-infinite
calculation). At low $\delta_0$, the surface tension $\sigma\dr{e}$ is
larger for the set RSk1* than for SkM*, while the contrary happens
with $\sigma_\mu$. This is so because of the values of $Q$ for both
interactions, as the LDM equations \req{eq49} and \req{eq50} show.
Since by construction both forces have the same $J$, the quantity $9
J^2 / 4 Q$ is larger for RSk1* and thus the initial slope of the
relativistic $\sigma\dr{e}$ and $\sigma_\mu$ is steeper than for SkM*.
Altogether, the above hints at a difference in the asymmetry
dependence of the nuclear surface in the relativistic model as
compared to conventional Skyrme forces.

To conclude this discussion we would like to clarify why the
relativistic parametrizations give larger values of $L$. This fact can
be traced back to the behaviour of the symmetry energy with density.
From Eq.\ \req{eq51} and Appendix A, for Skyrme forces it is
%
%Equation 550
\beqa
 J(\rho) & = & \frac{\hbar^2 k\dr{F}^2}{6 m}
- \frac{1}{4} t_0 \lf( \frac{1}{2}+x_0 \ri) \! \rho
+  \frac{1}{24} \lf( \frac{3 \pi^2}{2} \ri)^{2/3}
 \lf[ t_2 (4+5 x_2) - 3 t_1 x_1 \ri] \rho^{5/3}
\nonumber \\[3mm]
& & \mbox{}
- \frac{1}{24} t_3 \lf( \frac{1}{2}+x_3 \ri) \rho^{1+\alpha} ,
\label{eq550} \eeqa
with $k_{\rm F} = (3 \pi^2 \rho/2 )^{1/3}$. The first term comes from
the kinetic energy and the other ones from the interaction. For the
considered forces the term with $t_0 \rho$ gives a positive
contribution, whereas the terms with $\rho^{5/3}$ and
$\rho^{1+\alpha}$ produce a negative contribution. The net result is
that the function $J(\rho)$ initially grows with density, but it is
progressively bent as the density increases. In the relativistic model
one has (Appendix A, $\hbar= c= 1$)
%
%Equation 551
\beq
 J(\rho) =  \frac{k\dr{F}^2}{6 \ep\dr{F}}
          + \frac{1}{8} \frac{g_\rho^2}{m_\rho^2} \rho \,.
\label{eq551} \eeq
The first quantity is the relativistic kinetic contribution, while the
isovector $\rho$ meson is the mechanism of the relativistic
interaction to introduce additional symmetry energy. In Eq.\
\req{eq551} there are no negative terms that oppose the growing
behaviour of $J(\rho)$. Since the coefficient $L$ is essentially the
derivative of $J(\rho)$ evaluated at $\rho\dr{nm}$, cf.\ Eq.\
\req{eq51}, this explains a higher $L$ in the relativistic model.

%
% >>>>>>>>>>>>>>>>>>>>>>>>>>>>>>>>>>>>>>>>>>>>>>>>>>>>>>>>>>>>>>>>>>>>
% Comparison with previous calculations.
%
\subsection{Comparison with previous calculations}
\hspace*{\parindent}
For the Skyrme forces, our results can be contrasted with those of
Kolehmainen {\it et al.}\ \cite{Kol85} (KPLT hereafter). Our
calculation differs from KPLT in two aspects. On the one hand, our
approach is fully variational whereas in KPLT the densities were
restricted to have the form of a trial Fermi function to a power (plus
a constant in the case of drip). On the other hand, we have kept the
coefficient of the Weizs\"acker term in the second-order energy
functional equal to its standard ETF value, namely $\beta=1/36$.
However, in several applications of KPLT it was set to $\beta=1/18$
which somehow simulates effects of order $\hbar^4$ \cite{Bra85,Tre86}.
From Table 2 one can see that, for $\beta=1/36$, the results obtained
in KPLT by means of a parametrized density are very close to those we
obtain with fully variational densities (see also Ref.\ \cite{Ce90a}).
It is also clear that the use of $\beta=1/18$ or $\beta=1/36$ has a
drastic effect on the surface properties.

For the SkM* and SIII interactions we may compare our results for the
surface tensions $\sigma\dr{e}$ and $\sigma_\mu$, shown in Fig.\ 12 as
a function of $\delta_0^2$, with the corresponding values presented in
Fig.\ 7 of KPLT (called $\sigma\dr{s}$ and $\omega\dr{s}$
respectively). The KPLT calculation was performed for $\beta=1/18$,
but we find a fair agreement between both calculations if our results
are scaled by a factor $\sigma\dr{KPLT,0}/\sigma_0$ (the ratio at
$\delta_0= 0$ of the KPLT value with $\beta=1/18$ to ours with
$\beta=1/36$, that can be read from Table 2).

From the comparison of Fig.\ 13 of this work with Fig.\ 8 of KPLT, we
see that the agreement is also good for the Gibbs curvature energy per
unit length $\gamma_\mu$ (called $\omega\dr{c}$ in KPLT), when our
results are scaled by $\gamma\dr{KPLT,0}/\gamma_0$ (it can be read
from Table 2). However, a remarkable discrepancy appears when the
values for $\gamma\dr{e}$ (called $\sigma\dr{c}$ in KPLT) are
compared. After scaling our values with $\gamma\dr{KPLT,0}/\gamma_0$,
they are larger than those reported in KPLT\@. The disagreement
persists if we perform our self-consistent calculation with $\beta=
1/18$. One must note that the authors of KPLT did not include the
dynamical curvature energy which comes from the {\em implicit}
curvature dependence of the nuclear densities (see Section 3 and
Appendix A of KPLT). This is a non-vanishing contribution in the case
of $\gamma\dr{e}$ at $\delta_0 \neq 0$, Eq.\ (\ref{eq46}).

In Table 2 we also compare our surface properties for the relativistic
model with those reported by Von-Eiff {\it et al.}\ \cite{Von94b}
(VPSW hereafter). The VPSW results were computed at the TF level,
without the gradient corrections we have in the RETF method. For the
surface energy coefficient, in the symmetric case, one finds lower
values in the RETF than in the TF calculation. As discussed in Refs.\
\cite{Von94a,Cen93b,Cen92,Spe93}, this is a typical feature of the
relativistic sets that have a bulk effective mass $m^*_\infty/m$,
roughly, below 0.70. From Table 2 we see that the surface thickness
$t$ turns out to be smaller in RETF than in TF, in accordance with the
trend of the surface energies. Finally, the surface stiffness $Q$ is
larger in the RETF calculation than in TF, which expresses a greater
rigidity of the nuclear system in RETF against the separation of the
neutron and proton surfaces.

%
% >>>>>>>>>>>>>>>>>>>>>>>>>>>>>>>>>>>>>>>>>>>>>>>>>>>>>>>>>>>>>>>>>>>>
% Nuclear matter and the surface in the relativistic model.
%
\subsection{Nuclear matter and the surface in the relativistic model}
\hspace*{\parindent}
Next, we analyze the dependence of the surface on the various
magnitudes that characterize the infinite nuclear matter in the
relativistic case. The saturation properties of nuclear matter are
governed by the meson coupling-to-mass ratios $g\dr{s}^2/m\dr{s}^2$,
$g\dr{v}^2/m\dr{v}^2$ and $g_\rho^2/m_\rho^2$, and by the non-linear
couplings $g_2$ and $g_3$ \cite{Ser86,Bog77} (see Appendix A).
Contrarily, the surface properties extracted from the semi-infinite
system depend on the meson coupling constants and masses separately.

We start by obtaining the surface properties with a non-linear \sw set
whose parameters we have fitted to these nuclear matter properties:
volume energy $a\dr{v}= -16$ MeV, saturation density $\rho\dr{nm}=
0.16$ fm$^{-3}$, incompressibility $K= 200$ MeV, effective mass
$m^*_\infty/m= 0.70$, and bulk symmetry energy $J= 30$ MeV\@. The
scalar meson mass is $m\dr{s}= 500$ MeV\@. We call this set of
parameters NLM\@. Later we recalculate the surface quantities changing
one of the properties that define the set NLM at each time, with the
other properties fixed to their initial values. This way we can study
the individual effect of each bulk property and of the scalar mass
$m\dr{s}$ on the surface. The results are collected in Table 3 for the
symmetric case $\delta_0= 0$, and in Table 4 for $\delta_0= 0.212$
(below neutron drip, corresponding to an ideal system of $^{208}$Pb)
and $\delta_0= 0.6$ (above neutron drip).

Table 3 also shows the value of the neutron excess in uniform matter
when drip neutrons appear ($\bar\delta\dr{nd}$) for NLM and the
related sets. It can be seen that important changes in
$\bar\delta\dr{nd}$ are connected with varying the volume and symmetry
energies. This is due to the fact that the neutron drip is mainly
determined by the ratio $a\dr{v}/J$. On the other hand, by decreasing
the effective mass or the incompressibility, the neutron drip point
moves to higher values of $\delta_0$. Of course, $\bar\delta\dr{nd}$
is not changed by variations of $m\dr{s}$ that are compensated by
changes in the scalar coupling constant $g\dr{s}$. The proton drip
point is less sensitive to the nuclear matter properties and appears
at $\delta_0 \sim 0.7$ for all the examined cases, and thus we do not
display it in Table 3.

Since the scalar mass sets the range of the attractive scalar
interaction, there is a direct correlation between $m\dr{s}$ and the
surface properties. A larger $m\dr{s}$ determines a shorter range of
the attractive potential. As Tables 3 and 4 show, this leads to a
steeper surface and to a visible reduction of the surface and
curvature energies and of the surface thickness $t$ and the neutron
skin thickness $\Theta$. The surface also is strongly correlated with
the value of the effective mass of the interaction. Indeed,
$m^*_\infty/m$ plays a prominent role in the majority of nuclear
structure properties in the relativistic theory, since it is
intimately related to the vector and scalar fields \cite{Rei89,Gam90}.
For the symmetric case and small asymmetries, on increasing
$m^*_\infty/m$ the surface and curvature energy coefficients and $t$
and $\Theta$ show a downward trend, as it happened with the scalar
mass. The same situation is found at $\delta_0= 0.6$ except for
$E\dr{s,\mu}$, which turns out to be larger for $m^*_\infty/m= 0.70$
than for $m^*_\infty/m= 0.55$. The effective mass provides a measure
of the non-local effects, which contribute to make the surface more
diffuse \cite{Fri75}. A higher effective mass is associated with less
non-locality and thus it tends to favour a sharper surface.

A smaller value of the incompressibility $K$ softens the nuclear
surface and the thickness $t$ augments, while the surface and
curvature energies decrease. The change of the neutron skin thickness
$\Theta$ induced by $K$ is not monotonous with $\delta_0$. From Tables
3 and 4 it can be checked that the main changes in the ratio
$E\dr{s}/t$ come from the value of $K$, as discussed in the literature
\cite{Bog77,Hof89}. Increasing $K$ brings about larger $E\dr{s}/t$
values. This ratio experiences a clear reduction for high $\delta_0$.
For instance, for the set NLM one has $E\dr{s}/t= 8.8$ MeV/fm at
$\delta_0= 0$, 9.1 (6.0) MeV/fm at $\delta_0= 0.212$, and 3.7 (0.4)
MeV/fm at $\delta_0= 0.6$ (the first number corresponds to
$E\dr{s,e}/t$ and the number in parentheses to $E\dr{s,\mu}/t$).

Modifying the saturation density $\rho\dr{nm}$ or the volume energy
$a\dr{v}$ has little consequences on the surface properties. The
analysis of Tables 3 and 4 shows that $\rho\dr{nm}$ has a moderate
effect on the surface and curvature energies, while the surface
thickness $t$ is not sensitive to $\rho\dr{nm}$. Conversely, the
incidence of $a\dr{v}$ is more visible on the surface thickness than
on the energies. The neutron skin thickness $\Theta$ is almost
unaffected by $\rho\dr{nm}$ and $a\dr{v}$.

The coefficient $J$ does not change the surface properties of the
symmetric system. The surface energy $E\dr{s,e}$ and the neutron skin
$\Theta$ increase with $J$ at $\delta_0= 0.212$, but the opposite
tendency is found at $\delta_0= 0.6$. At both values of $\delta_0$,
$E\dr{c,e}$ and the thickness $t$ become larger when $J$ is raised,
whereas $E\dr{s,\mu}$ and $E\dr{c,\mu}$ decrease. Reading from Table 3
the value of $Q$ for NLM and for the set with $J= 40$ MeV, the
behaviour of $\Theta$, $E\dr{s,e}$ and $E\dr{s,\mu}$ with $J$ at
$\delta_0= 0.212$ is consistent with the LDM equations
\req{eq48}--\req{eq50}.

We can see from Table 3 that the surface-stiffness coefficient $Q$ is
raised by larger values of $m\dr{s}$, $m^*_\infty/m$ and $J$. This
behaviour of $Q$ with $m\dr{s}$ and $m^*_\infty/m$ is not
surprising because the surface becomes stiffer when these quantities
increase, and it is harder to separate the neutron and proton
surfaces. Furthermore, for small $\delta_0$ at least, it reflects the
inverse proportionality between $Q$ and the neutron skin $\Theta$
shown by Eq.\ \req{eq48}, since $\Theta$ decreases with $m\dr{s}$ and
$m^*_\infty/m$ (at fixed $\delta_0$). Reducing $\rho\dr{nm}$ and
$a\dr{v}$ or a greater $K$ also makes $Q$ grow, but the changes are
less noticeable.

In contrast with our finding that $Q$ increases with $J$ when the
other properties of the force are not changed, in previous literature
increasing values of $J$ have been related with decreasing values of
$Q$ \cite{Kol85,Von94b}. Certainly, this is true if the total
(bulk-plus-surface) symmetry energy $E\dr{sym}$ of the interaction is
set to the empirical value fitted by mass formulae, that contain a
term of the type $a\dr{sym}(N-Z)^2/A$. Introducing $I= (N-Z)/A$, in
the LDM one has \cite{Mye69}
%
%Equation 53
\beq
 E\dr{sym} = a\dr{sym} \, I^2 A =
 \lf[ J - \lf( \frac{9J^2}{4Q}
 - \frac{2 E\dr{s,0} L}{K} \ri) A^{-1/3} \ri] I^2 A \,,
\label{eq53} \eeq
which is valid for large $A$ and small $I$. Equation \req{eq53} shows
that if the symmetry energy $E\dr{sym}$ of a given interaction is
fixed empirically, a large $J$ coefficient must be associated with a
small $Q$ coefficient and viceversa. On the contrary, if the
parameters of the interaction are not fitted to reproduce $E\dr{sym}$
such constraint between $J$ and $Q$ needs not be satisfied, as happens
e.g.\ when the parameters are chosen to give the nuclear matter
properties we wish.

To clarify this point we plot in Figs.\ 16 and 17, respectively, the
surface-stiffness coefficient $Q$ and the mass-formula symmetry
coefficient $a\dr{sym}$ (calculated for $^{208}$Pb) as a function of
$J$. They are drawn for several \sw and Skyrme parametrizations; we
have added the relativistic set SRK3M5 ($J= 23.5$ MeV) \cite{Cen92}
and the Skyrme force SI$^\prime$ ($J= 29.35$ MeV) \cite{Kol85,Lat81}
to the interactions already considered. It can be seen that forces
fulfilling that $Q$ decreases as $J$ increases (namely SIII,
SI$^\prime$, SkM*, NL-SH and NL1) roughly lie on a curve in the $Q-J$
plane (Fig.\ 16). These same forces lie in a rather narrow region of
values of $a\dr{sym}$ (22--24 MeV, the empirical region) in the
$a\dr{sym}-J$ plane (Fig.\ 17). There are two forces, namely NL2 and
SRK3M5, that clearly deviate from the general tendency in the $Q-J$
plane. The reason for this anomaly is that these two forces are
clearly outside the empirical region where the remaining forces lie in
the $a\dr{sym}-J$ plane.

For further insight we have drawn in Figs.\ 16 and 17 the results of
two more forces, NL1J4 and NL1J5. They are identical to the set NL1
excepting the value of the bulk symmetry energy: $J= 40$ MeV for NL1J4
and $J= 50$ MeV for NL1J5. The coefficient $a\dr{sym}$ for NL1J4 and
NL1J5 moves away from the empirical band, but both $Q$ and $a\dr{sym}$
grow with $J$ because the remaining nuclear matter properties have
been kept constant. Thus, to get the right symmetry coefficient
$a\dr{sym}$ one must change various properties of the effective
interaction simultaneously, and then the tendency of smaller $Q$ with
higher $J$ is fulfilled.

%
% >>>>>>>>>>>>>>>>>>>>>>>>>>>>>>>>>>>>>>>>>>>>>>>>>>>>>>>>>>>>>>>>>>>>
% From semi-infinite matter to finite nuclei.
%
\subsection{From semi-infinite matter to finite nuclei}
\hspace*{\parindent}
To conclude this section, it may be interesting to compute the energy
of large asymmetric and uncharged nuclei using the mass-formula
coefficients calculated in the semi-infinite medium. We can then check
the results against those we obtain from a self-consistent,
semiclassical calculation of spherical finite nuclei
\cite{Cen93b,Cen92,Ce90a} without Coulomb interaction. Before
presenting the numbers we pass to discuss some aspects referred to the
mass formulae we will utilize.

In writing the energy of a nucleus one has two choices, according to
the two definitions of the reference energy, Eqs.\ \req{eq26} and
\req{eq31}. One possibility is
%
%Equation 4.8
\beq
E = E\dr{v,e}(\delta_0) \, A + E\dr{s,e}(\delta_0) \, A^{2/3}
    + E\dr{c,e}(\delta_0) \, A^{1/3} \,,
\label{eq4.8} \eeq
with an e-volume energy $E\dr{v,e} = (\ed_0 - \ed\dr{d})/(\rho_0 -
\rho\dr{d})$, see Eq.\ \req{eq26}. The other alternative to write the
energy is
%
%Equation 4.9
\beq
E = E\dr{v,\mu}(\delta_0) \, A + E\dr{s,\mu}(\delta_0) \, A^{2/3}
    + E\dr{c,\mu}(\delta_0) A^{1/3} \,,
\label{eq4.9} \eeq
in terms of a Gibbs volume energy $E\dr{v,\mu}= \mu\dr{n}(N -
N\dr{d})/A + \mu\dr{p}(Z -Z\dr{d})/A$, see Eq.\ \req{eq31}. The
surface energy coefficients $E\dr{s,e}$ and $E\dr{s,\mu}$ and the
curvature energy coefficients $E\dr{c,e}$ and $E\dr{c,\mu}$ have been
defined in Eqs.\ \req{eq33}, \req{eq34} and \req{eq39}.

In order to establish a link with the usual mass formulae we shall
restrict ourselves to values of $\delta_0$ below the neutron drip
point. In this case $E\dr{v,e}(\delta_0)$ is just the energy per
particle in bulk matter $\ed/\rho$, calculated at the saturation
density $\rho\dr{sat}$ for $\delta_0$. Likewise, in
$E\dr{v,\mu}(\delta_0)$ the neutron and proton chemical potentials
will be evaluated at $\rho\dr{sat}$ and $\delta_0$, with $N\dr{d}=
Z\dr{d}= 0$.

When one deals with finite nuclei it must be taken into account that
the overall neutron excess $I$ of the system,
%
%Equation 4.10
\beq
I = \frac{N-Z}{A} \,,
\label{eq4.10} \eeq
is in general different from the bulk neutron excess $\delta_0$. Using
Eqs.\ \req{eq3} and \req{eq4} the difference between the total volume
energies reads
%
%Equation 4.11
\beqa
E\dr{v,\mu} (\delta_0) A - E\dr{v,e}(\delta_0) A & = &
\lf( \frac{\partial\ed}{\partial \rho} +
     \frac{\partial \ed}{\partial \delta_0}
     \frac{1 - \delta_0}{\rho} \ri) N
+ \lf( \frac{\partial \ed}{\partial \rho} -
       \frac{\partial \ed}{\partial \delta_0}
       \frac{1 + \delta_0}{\rho} \ri) Z - \frac{\ed}{\rho} A
\nonumber \\[3mm]
& = &
 \frac{1}{\rho} \frac{\partial \ed}{\partial \delta_0}
 (I - \delta_0) A \,,
\label{eq4.11} \eeqa
with all variables evaluated at the density $\rho\dr{sat}$ for
$\delta_0$. The Taylor expansion of $E\dr{v,e}(I)$ about $\delta_0$
gives
%
%Equation 4.11b
\beq
E\dr{v,e}(I) = E\dr{v,e}(\delta_0)
+ \frac{1}{\rho} \frac{\partial \ed}{\partial \delta_0}
  \, (I - \delta_0)
+ \frac{1}{2 \rho} \frac{\partial^2 \ed}{\partial \delta_0^2}
  \, (I - \delta_0)^2 + \ldots \,.
\label{eq4.11b} \eeq
Comparison of Eq.\ \req{eq4.11b} with Eq.\ \req{eq4.11} shows that
$E\dr{v,\mu}(\delta_0) - E\dr{v,e}(I)$ is of second order in the small
quantity $I- \delta_0$ and, consequently, one can approach
$E\dr{v,\mu}(\delta_0)$ by $E\dr{v,e}(I)$. Since in the limit $I \to
0$ (or $\delta_0 \to 0$) it is $E\dr{v,e}(I) = a\dr{v} + J I^2$, for
small $\delta_0$ and $I$ we can write
%
%Equation 4.12
\beq
 E\dr{v,\mu}(\delta_0) = a\dr{v} + J I^2 .
\label{eq4.12} \eeq

Equation \req{eq4.11} is actually a surface term (i.e., proportional
to $A^{2/3}$). This can be seen as follows. Defining the effective
neutron and proton volumes as \cite{Mye85}
%
%Equation 4.14
\beq
V\dr{n} = \frac{N}{\rho\dr{n}}\hspace{1cm} \mbox{and} \hspace*{1cm}
V\dr{p} = \frac{Z}{\rho\dr{p}} \,,
\label{eq4.14} \eeq
one finds
%
%Equation 4.15
\beq
\frac{\Delta V}{V} = \frac{V\dr{n} -V\dr{p}}{V} =
\frac{N/\rho\dr{n} - Z/\rho\dr{p}}{A/\rho} =
\frac{2 (I - \delta_0)}{1 - \delta_0^2} .
\label{eq4.15} \eeq
In the limit of a very large nucleus $\Delta V/V$ can be written as
$\Theta S/ V$ \cite{Mye85}, where $S$ is the surface area and $\Theta$
the neutron skin thickness. With $V= 4 \pi R^3/3$ and $S= 4 \pi R^2$,
%
%Equation 4.16
\beq
\frac{\Delta V}{V} =
\frac{3 \Theta}{R} = \frac{3 \Theta}{r_0} A^{-1/3} .
\label{eq4.16} \eeq
Combining \req{eq4.15} with \req{eq4.16} and neglecting $\delta_0^2$
in front of unity leads to
%
%Equation 4.17
\beq
I - \delta_0 \simeq \frac{3 \Theta}{2 r_0} A^{-1/3} ,
\label{eq4.17} \eeq
which confirms the statement that \req{eq4.11} is proportional to
$A^{2/3}$.

For small $\delta_0$, and recalling Eq.\ \req{eq48} for $\Theta$, from
Eq.\ \req{eq4.17} it is easy to show that the bulk neutron excess
$\delta_0$ and the overall neutron excess $I$ are related by
%
%Equation 4.18
\beq
\delta_0 \simeq I \lf( 1 + \frac{9J}{4Q}A^{-1/3} \ri)^{-1}
 \simeq I \lf( 1 - \frac{9J}{4Q} A^{-1/3} \ri) ,
\label{eq4.18} \eeq
provided that $(9J/4Q) A^{-1/3} \ll 1$. This is an interesting result,
since it represents a new manner to compute the surface-stiffness
coefficient $Q$, in addition to Eqs.\ \req{eq48} and \req{eq52}. By
calculating a series of large and uncharged finite nuclei with $I$
constant, and evaluating $\delta_0$ from the central variational
densities, from Eq.\ \req{eq4.18} one can extract the value of $Q$ for
the considered interaction.

Using the LDM expansion \req{eq50} for $E\dr{s,\mu}(\delta_0)$ plus
Eq.\ \req{eq4.18}, and neglecting asymmetry effects in the curvature
energy ($I^2 A^{1/3}$ terms), one recovers from Eq.\ \req{eq4.9} the
liquid droplet mass formula for an uncharged nucleus of small
asymmetry \cite{Mye69}:
%
%Equation 4.13
\beqa
E & = & a\dr{v}A + E\dr{s,0} A^{2/3}
+ \lf( E\dr{c,0} - \frac{2E\dr{s,0}^2}{K} \ri) A^{1/3}
\nonumber \\[3mm]
& & \mbox{}
+ \lf[ J - \lf(\frac{9J^2}{4Q} - \frac{2E\dr{s,0}L}{K} \ri) A^{-1/3}
\ri] I^2 A \,.
\label{eq4.13} \eeqa
In the above equation $E\dr{s,0}$ and $E\dr{c,0}$ are the surface and
curvature energy coefficients for the symmetric case, and the
corrective term $-(2E\dr{s,0}^2/K) A^{1/3}$ accounts for the surface
compression effects that appear in finite nuclei \cite{Mye69,Sto73}.
The formula \req{eq4.13} can also be obtained from Eq.\ \req{eq4.8} by
expanding $E\dr{v,e}(\delta_0)$ as $a\dr{v} + J \delta_0^2$ and using
Eq.\ \req{eq4.18} together with the expansion \req{eq49} for
$E\dr{s,e} (\delta_0)$.

Tables 5 and 6 collect the energy per nucleon, obtained in various
ways, for uncharged large nuclei using the SkM* and NL2 interactions.
The mass number ranges from $A= 250$ to $A= 20000$, while the overall
neutron excess is fixed at $I= 0.2$. In both tables we list in the
second column $\delta_0(A)$, which we obtain from the interior
densities produced by the self-consistent calculation of the finite
nucleus with $A$ particles:
%
%Equation 4.19
\beq
\delta_0 =
\frac{\rho\dr{n}(0) - \rho\dr{p}(0)}{\rho\dr{n}(0) + \rho\dr{p}(0)}
\,.
\label{eq4.19} \eeq
Note from the tables that $\delta_0$ approaches $I$ as $A$ grows. The
energies (per nucleon) of the finite nuclei are given in the third
column of Tables 5 and 6, labeled by $E\dr{FN}(I)$. In the next two
columns we show the predictions of the mass formulae \req{eq4.8} and
\req{eq4.9}, labeled by $E\dr{MF,e}(\delta_0)$ and
$E\dr{MF,\mu}(\delta_0)$ respectively. In both cases we have included
an additional term $-(2E\dr{s,0}^2/K)$ to account for the
compressional effect. The rightmost column $E\dr{LDM}(I)$ exhibits the
output of the LDM mass formula \req{eq4.13}.

The $E\dr{MF,\mu}(\delta_0)$ results agree almost perfectly with the
self-consistent values $E\dr{FN}(I)$ for the largest analyzed nuclei.
The differences to $E\dr{FN}(I)$ are in all cases smaller than 0.5\%,
a similar quality to that found in Ref.\ \cite{Cen93c} for a
calculation performed in the symmetric case. Though the quality of
$E\dr{MF,e}(\delta_0)$ is also remarkably good, the overall agreement
with the self-consistent calculation is a little worse. It was
discussed in Section 3 that the surface tension which is minimized in
the semi-infinite calculation is the Gibbs surface tension
$\sigma_\mu$, while $\sigma\dr{e}$ is not. The liquid droplet model
results also agree very well with $E\dr{FN}(I)$, and by extension with
$E\dr{MF,\mu}(\delta_0)$ and $E\dr{MF,e}(\delta_0)$, as expected from
the small value of $I$. This fact shows that Eqs.\ \req{eq4.8} and
\req{eq4.9} have the right limit for small $\delta_0$.

With respect to the LDM mass formula \req{eq4.13}, Eqs.\ \req{eq4.8}
and \req{eq4.9} show two main differences. On the one hand, the LDM
volume and surface energy coefficients have been expanded up to
quadratic terms in asymmetry, while our calculation includes it to all
orders. On the other hand, the LDM expression neglects the dependence
on asymmetry of the curvature energy term, which we have taken into
account in Eqs.\ \req{eq4.8} and \req{eq4.9}. Looking at each
coefficient separately, we have checked that the relative difference
(between our calculation and LDM) is much more important in the
curvature energy coefficient than in the other ones. However, due to
the $A$ dependence of the mass formula, for very large nuclei most of
the final discrepancy between our total energies and the LDM ones
comes from the volume and surface terms. This is no longer the case
for small nuclei, where the curvature term is mainly responsible for
the disagreement.

An advantageous feature of the mass formulae \req{eq4.8} and
\req{eq4.9} with respect to the LDM one is that they also could be
employed when drip particles exist. In such a case, one should use the
expressions of the \mbox{e-} and $\mu$-volume energies ($E\dr{v,e}$
and $E\dr{v,\mu}$) with drip. The compressional energy should be
calculated following techniques similar to those of Ref.\
\cite{Gui95}, where this correction was derived at finite temperature
for helium clusters.

The neutron skin thickness $\Theta$ can be extracted from the finite
nuclei results through Eq.\ \req{eq4.17}. Reading $\delta_0$ from
Tables 5 and 6, we have found a good agreement with reference to the
values of $\Theta$ calculated in the semi-infinite medium, specially
for large $A$. The agreement worsens slightly if one defines the
neutron skin of a finite nucleus to be (see Ref.\ \cite{Tre86} for a
discussion)
%
%Equation 4.20
\beq
 \Theta= R\dr{n} - R\dr{p} \,,
\label{eq4.20} \eeq
replacing the actual neutron and proton distributions by spheres of
radii $R\dr{n}$ and $R\dr{p}$ with constant densities $\rho\dr{n}(0)$
and $\rho\dr{p}(0)$:
%
%Equation 4.21
\beqa
 \frac{4}{3} \pi R\dr{n}^3 \, \rho\dr{n}(0) & = & N
\nonumber \\[3mm]
 \frac{4}{3} \pi R\dr{p}^3 \, \rho\dr{p}(0) & = & Z \,.
\label{eq4.21} \eeqa
It can be shown that Eq.\ \req{eq4.20} transforms into Eq.\
\req{eq4.17} in the the limit of small asymmetries and to first order
in $A^{1/3}$.

As told before, one can calculate $Q$ from the difference $I-
\delta_0$ by means of Eq.\ \req{eq4.18}. From the values of Tables 5
and 6 we obtain $Q= 39.5$ MeV for SkM* and $Q= 41.9$ MeV for NL2.
These results agree nicely with those reported in Table 2 that were
obtained directly from the semi-infinite nuclear matter calculations,
Eqs.\ \req{eq48} and \req{eq52}.

\pagebreak
% >>>>>>>>>>>>>>>>>>>>>>>>>>>>>>>>>>>>>>>>>>>>>>>>>>>>>>>>>>>>>>>>>>>>
% SUMMARY AND CONCLUDING REMARKS.
%
\section{Summary}
\hspace*{\parindent}
In this paper we have investigated the surface properties of
asymmetric semi-infinite nuclear matter with arbitrary neutron excess.
This has been done within a semiclassical context, by means of density
functional techniques, for non-relativistic and relativistic models.
Specifically, we have used the extended Thomas--Fermi approach
including gradient corrections of order $\hbar^2$, together with
Skyrme forces in the non-relativistic case and the non-linear \sw
model in the relativistic case.

First, we have discussed the coexistence between a nucleus and drip
particles under the bulk equilibrium approximation. Next, we have
studied the surface properties of two-component systems. We have found
the neutron and proton density profiles in the semi-infinite geometry
by solving self-consistently the variational Euler--Lagrange
equations, at given values of the bulk neutron excess. General trends
of the evolution of the nuclear surface with the asymmetry have been
obtained by exploring the surface thickness $t$, the surface width $b$
and the neutron skin thickness $\Theta$.

We have treated the calculation of the surface and curvature energy
coefficients according to the two definitions of the reference energy
$E\dr{ref}$. The self-consistent calculation of the density profiles
corresponds to the minimization of the surface tension with the Gibbs
prescription for $E\dr{ref}$. An important question is the separation
of the curvature energy into geometrical and dynamical parts. Though
in the Gibbs prescription one can avoid the dynamical term by partial
integrations of the Laplacian, there exists always a non-vanishing
dynamical contribution at $\delta_0 \neq 0$ if the e-definition of
$E\dr{ref}$ is chosen. We also have calculated the surface stiffness
coefficient $Q$. We have extracted it from the neutron skin $\Theta$
and from the surface energy coefficients, finding a good agreement
between both methods.

To ascertain the origin of some differences in the surface properties
between the non-relativistic and relativistic models, we have built up
a non-linear \sw parametrization with the same nuclear matter
properties and surface energy at zero asymmetry as SkM*. The
non-equivalent behaviour in the evolution of the surface with
asymmetry is mainly due to the fact that the surface-stiffness
coefficient is smaller for the relativistic parametrization.

Comparing our results for Skyrme forces with those of Kolehmainen {\it
et al.}\ \cite{Kol85} we find a good correspondence, except for the
curvature energy per unit length $\gamma\dr{e}$ for which the implicit
curvature dependence was not considered in Ref.\ \cite{Kol85}. In the
relativistic model we could make some comparisons with the
Thomas--Fermi results of Von-Eiff {\it et al.}\ \cite{Von94b}. For
interactions like NL1 and NL-SH with a small effective mass, including
the $\hbar^2$ inhomogeneity corrections reduces the surface energy and
thickness, while the surface stiffness $Q$ increases.

Our work has also been concerned with the analysis of the impact on
the surface-symmetry properties of the various quantities that
characterize the relativistic interaction. The strongest dependences
have been found with the scalar mass $m\dr{s}$, that determines the
range of the attractive potential, and with the effective mass
$m_\infty^*/m$, that somehow reflects the non-local effects.

The surface and curvature coefficients derived in this paper allow to
write mass formulae that can be extended to the case when drip
particles exist. They could be very useful in physical situations
involving large asymmetries. We have checked the predictions of our
mass formulae with the self-consistent energies of a calculation of
uncharged large finite nuclei, and with the LDM mass formula in the
low asymmetry limit. The agreement with the self-consistent
calculations is as good as in the symmetric case.

Hot asymmetric nuclear matter and finite nuclei have been investigated
in non-relativistic and relativistic calculations. However, to our
knowledge, hot semi-infinite nuclear matter has been analyzed only for
the symmetric case in the relativistic mean field theory \cite{Mul93}.
A natural extension of our work would be to study the surface and
curvature properties of asymmetric semi-infinite nuclear matter at
finite temperature, since the combined effect of asymmetry and
temperature has an important bearing on astrophysical objects and
energetic heavy ion collisions.

% >>>>>>>>>>>>>>>>>>>>>>>>>>>>>>>>>>>>>>>>>>>>>>>>>>>>>>>>>>>>>>>>>>>>
% ACKNOWLEDGEMENTS
%
\section*{Acknowledgements}
\hspace*{\parindent}
The authors would like to acknowledge support from the DGICYT (Spain)
under grant PB95-1249 and from the DGR (Catalonia) under grant
GR94-1022. M. Del Estal acknowledges in addition financial support
from the CIRIT (Catalonia).

\pagebreak
% >>>>>>>>>>>>>>>>>>>>>>>>>>>>>>>>>>>>>>>>>>>>>>>>>>>>>>>>>>>>>>>>>>>>
\section{Appendix A}
\hspace*{\parindent}
In the non-relativistic case with Skyrme forces the energy density for
an uncharged nucleus can be written as \cite{Bra85,Vau72}
%Equation A1
\beqa
 \ed & = &\frac{\hbar^{2}}{2m} (f\dr{n} \tau\dr{n} +
f\dr{p} \tau\dr{p}) +
\frac{1}{2}
t_0 \lf[\lf(1 + \frac{x_0}{2}\ri)\rho^{2} - \lf(x_{0} + \frac{1}{2}
\ri)(\rho\dr{n}^{2} + \rho\dr{p}^{2})\ri]
\nonumber \\[3mm]
&  & \mbox{}
- \frac{1}{16}\lf[t_2  \lf( 1  + \frac{x_{2}}{2}\ri)- 3t_1 \lf(1 +
\frac{x_{1}}{2}\ri) \ri](\mbox{\boldmath $\nabla$} \rho)^2
\nonumber \\[3mm]
& & \mbox{}
- \frac{1}{16}\lf[3t_1 \lf( x_1 + \frac{1}{2}\ri) + t_2 \lf(x_2 +
\frac{1}{2}\ri) \ri] \lf[ (\mbox{\boldmath $\nabla$}\rho\dr{n})^2 +
(\mbox{\boldmath $\nabla$}\rho\dr{p})^2\ri]
\nonumber \\[3mm]
& & \mbox{}
+ \frac{1}{12} t_3 \rho^\alpha
 \lf[ \lf( 1 + \frac{x_{3}}{2}\ri)\rho^2- \lf(x_3 +
 \frac{1}{2}\ri)(\rho\dr{n}^2+\rho\dr{p}^2) \ri]
\nonumber \\[3mm]
& & \mbox{}
- \frac{1}{2}W_0 \lf(\rho \mbox{\boldmath $\nabla$} \scdot {\bf J}
+ \rho\dr{n}\mbox{\boldmath $\nabla$}
\scdot {\bf J}\dr{n} + \rho\dr{p}
 \mbox{\boldmath $\nabla$} \scdot {\bf J}\dr{p} \ri) .
\label{eqA1} \eeqa
The quantity $f_{q}$ ($q$ = n,p) is related with the effective mass
of each kind of nucleon through
%Equation A2
\beqa
f_{q}  =  \frac{m}{m_{q}^{*}} & = & \frac{2m}{\hbar^2} \frac{\partial
\ed}{\partial
\tau_{q}} = 1  + \frac{m}{2\hbar^2}\lf\{\lf[t_1\lf(1 +
\frac{x_1}{2}\ri) + t_2\lf(1+ \frac{x_2}{2}\ri)\ri]\rho\ri.
\nonumber \\[3mm]
& & \mbox{}
\lf. + \lf[t_2\lf(x_2 + \frac{1}{2}\ri) - t_1
  \lf(x_1+ \frac{1}{2}\ri)\ri]\rho_{q}\ri\} ,
\label{eqA2} \eeqa
where $\rho=\rho\dr{n} + \rho\dr{p}$ is the total particle density.

In the ETF approach to order $\hbar^2$, the kinetic energy density
$f_q \tau_q$ and the spin density ${\bf J}_{q}$ are given by
\cite{Gra79,Bra85}
%
%Equation A3
\beq
f_q \tau_q = \frac{3}{5}(3\pi^2)^{2/3}f_{q}\rho_{q}^{5/3} +
\frac{1}{36}f_{q}\frac{(\mbox{\boldmath $\nabla$}
 \rho_q)^2}{\rho_q} - \frac{1}{3} \mbox{\boldmath $\nabla$}f_q
\scdot {\mbox{\boldmath $\nabla$}\rho_{q}}
 - \frac{1}{12} \rho_q \frac{(\mbox{\boldmath $\nabla$}
f_{q})^2}{f_{q}} +
2\rho_{q} \frac{({\bf S}_{q})^2}{f_{q}}
\label{eqA3} \eeq
%Equation A4
\beq
{\bf J}_{q} = -\frac{2\rho_{q}}{f_{q}}{\bf S}_{q} \,,
\label{eqA4} \eeq
with
%Equation A5
\beq
{\bf S}_{q} = \frac{m}{2 \hbar^2} W_0 \lf(\mbox{\boldmath
$\nabla$}\rho + \mbox{\boldmath
$\nabla$}\rho_{q}\ri) \,.
\label{eqA5} \eeq

In the relativistic formulation, the mean field Hartree energy density
for an uncharged nucleus within the non-linear \sw model reads
\cite{Ser86,Rei86,Gam90}
%Equation A6
\beq
\ed = \sum_\alpha \varphi_\alpha^{\dagger}
\lf[ - i \mbox{\boldmath $\alpha$} \scdot \mbox{\boldmath $\nabla$}
+ \beta m^* + g\dr{v}V + \frac{1}{2} g_\rho \tau_3 R \ri]
 \varphi_\alpha + \ed\dr{f} \,,
\label{eqA6} \eeq
where $\tau_3$ is the third component of the isospin operator and the
subindex $\alpha$ runs over occupied shell-model orbitals of the
positive energy spectrum. The relativistic effective mass (or Dirac
mass) is defined by $m^* = m - g\dr{s}\phi$. $\ed\dr{f}$ represents
the additional contribution to the energy density coming from the
fields $\phi$, $V$ and $R$ associated with the $\sigma$, $\omega$ and
$\rho$ mesons respectively:
%
%Equation A7
\beqa
\ed\dr{f} & = & \frac{1}{2} \lf[ (\mbox{\boldmath $\nabla$} \phi )^2
 + m\dr{s}^2 \phi^2
\ri]
- \frac{1}{2} \lf[ (\mbox {\boldmath $\nabla$} V )^2
+ m\dr{v}^2 V^2 \ri]
- \frac{1}{2} \lf[ (\mbox{\boldmath $\nabla$} R )^2
+ m_\rho^2 R^2 \ri]
\nonumber \\[3mm]
 & & \mbox{}
+ \frac{1}{3} g_2 \phi^3 + \frac{1}{4} g_3 \phi^4 \,.
\label{eqA7} \eeqa
In the relativistic expressions we take units $\hbar= c = 1$.

The corresponding semiclassical energy density has a similar structure
to Eq.\ \req{eqA6}, except that the nucleon variables are the neutron
and proton densities instead of the wave functions. In the RETF
approach it reads \cite{Cen93b,Cen92}
%
%Equation A8
\beqa
\ed & = & \sum_{q} \frac{1}{8 \pi^2} \lf[ k_{{\rm F}q}
\ep{_{{\rm F}q}^{3}} + k_{{\rm F}q}^3 \ep_{{\rm F}q}
- {m^*}^4 \ln \frac{ k_{{\rm F}q} + \ep_{{\rm F}q} }{m^*} \ri]
\nonumber \\[3mm]
& & \mbox{}
+ \sum_{q} \lf[ B_{1q}(k_{{\rm F}q}, m^*)
 (\mbox{\boldmath $\nabla$} \rho_{q})^2
+ B_{2q} (k_{{\rm F}q}, m^*)\lf( \mbox{\boldmath $\nabla$}
 \rho_{q} \scdot \mbox{\boldmath $\nabla$} m^* \ri) \ri.
\nonumber \\[3mm]
& & \mbox{}
\lf. + B_{3q}(k_{{\rm F}q}, m^*)
 (\mbox{\boldmath $\nabla$} m^*)^2 \ri] + g\dr{v} V \rho
+ \frac{1}{2} g_\rho R (\rho\dr{p} - \rho\dr{n}) + \ed\dr{f} \,.
\label{eqA8} \eeqa
Here $k_{{\rm F}q} = (3 \pi^2 \rho_{q} )^{1/3}$ is the Fermi momentum
and $\ep_{{\rm F}q} = \sqrt{k_{{\rm F}q}^2 + {m^*}^2}$. The functions
$B_{iq}$ are the coefficients of the relativistic corrections of order
$\hbar^2$ to the TF model:
%
%Equation A9
\beqa
B_{1q}(k_{{\rm F}q}, m^*) & = &\frac{\pi^2}{24 k_{{\rm F}q}^3
\ep_{{\rm F}q}^2}
\lf( \ep_{{\rm F}q} + 2k_{{\rm F}q} \ln \frac{k_{{\rm F}q} +
\ep_{{\rm F}q}}{m^*} \ri)
\\[3mm]
%
%Equation A10
B_{2q}(k_{{\rm F}q}, m^*) & = &
\frac{m^*}{6 k_{{\rm F}q} \ep_{{\rm F}q}^2}
\ln \frac{k_{{\rm F}q} + \ep_{{\rm F}q}}{m^*}
\\[3mm]
%
%Equation A11
B_{3q}(k_{{\rm F}q}, m^*) & = &
\frac{k_{{\rm F}q}^2}{24 \pi^2 \ep_{{\rm F}q}^2}
\lf[\frac {\ep_{{\rm F}q}}{k_{{\rm F}q}}
- \lf( 2 + \frac{\ep_{{\rm F}q}^2}{k_{{\rm F}q}^2} \ri)
\ln \frac{k_{{\rm F}q} + \ep_{{\rm F}q}}{m^*} \ri] .
\label{eqA11} \eeqa
We remark that the semiclassical functionals \req{eqA3} and \req{eqA8}
do not contain any laplacian operators $\Delta$ because they have been
removed by suitable partial integrations.

\pagebreak
% >>>>>>>>>>>>>>>>>>>>>>>>>>>>>>>>>>>>>>>>>>>>>>>>>>>>>>>>>>>>>>>>>>>>
\section{Appendix B}
\hspace*{\parindent}
The expression \req{eq46} for $E\dr{c,e}\ur{dyn}$, the dynamical part
of the e-curvature energy coefficient, requires the evaluation of the
derivatives $\partial\rho_{q}/\partial \kappa$ (with $\kappa$ the
curvature). We summarize in this appendix how this is achieved. For
further details (but restricted to the symmetric problem) we refer the
reader to the works \cite{Dur93} and \cite{Cen96}. The starting point
is the Wigner--Kirkwood (WK) expansion of the particle density to
second order in $\hbar$. In the non-relativistic case it reads
\cite{Gra79,Rin80}
%
%Equation B1
\beqa
\rho_{q,\,{\rm nr}}\ur{WK} & = &
 \frac{1}{3\pi^2} \lf(\frac{2m}{\hbar^2}\ri)^{3/2}
\lf\{ \lf(\frac{\mu_q-V_q}{f_q}\ri)^{3/2}
+ \frac{\hbar^2}{16m} \lf[ \lf( \frac{7}{4}
  \frac{(\mbox{\boldmath $\nabla$} f_q)^2}{f_q^2}
 - 5\frac{\Delta f_{q}}{f_{q}} \ri)
 \lf(\frac{\mu_q-V_{q}}{f_{q}} \ri)^{1/2} \ri. \ri.
\nonumber \\[3mm]
& & \lf. \lf. \mbox{}
+ \lf( \frac{1}{2} \frac{ ( \mbox{\boldmath $\nabla$} f_{q} \scdot
  \mbox{\boldmath $\nabla$} V_{q} ) }{f_q^2}
 - \frac{\Delta V_{q}}{f_{q}} \ri)
 \lf(\frac{\mu_q-V_{q}}{f_{q}} \ri)^{-1/2} \ri. \ri.
\nonumber \\[3mm]
& & \lf. \lf. \mbox{}
 - \frac{1}{4} \frac{( \mbox{\boldmath $\nabla$} V_q)^2}{f_q^2}
 \lf(\frac{\mu_q-V_{q}}{f_{q}} \ri)^{-3/2} \ri] \ri\} ,
\label{eqB1} \eeqa
where $V_{q} = \delta {\cal{H}} / \delta \rho_{q}$ is the one-body
potential for a Hamiltonian $\cal{H}$.

In the relativistic model we have \cite{Cen93b}
%
%Equation B2
\beqa
\rho_{q,\,{\rm rel}}\ur{WK} & = &
\frac{k_{{\rm{F}}q}^3}{3\pi^2} + \frac{1}{24 \pi^2}
\lf[ \frac{1}{k_{{\rm{F}}q}}\lf( 3 -
\frac{\ep_{{\rm{F}}q}^2}{k_{{\rm{F}}q}^2}\ri)
(\mbox{\boldmath $\nabla$} V_q)^2
- \lf( 2 \frac{\ep_{{\rm{F}}q}}{k_{{\rm{F}}q}}
+ 4\ln \frac{k_{{\rm{F}}q}
+ \ep_{{\rm{F}}q}}{m^*} \ri)\Delta V_q \ri.
\nonumber \\[3mm]
& & \mbox{}
+ 2 \frac{\ep_{{\rm{F}}q}}{k_{{\rm{F}}q}m^*}\lf(3 -
\frac{\ep_{{\rm{F}}q}^2} {k_{{\rm{F}}q}^2}\ri) ( \mbox{\boldmath
$\nabla$}V_q \scdot \mbox{\boldmath $\nabla$}m^* )     +
\frac{1}{k_{{\rm{F}}q}}\lf(2
-\frac{\ep_{{\rm{F}}q}^2}{k_{{\rm{F}}q}^2}\ri)(
\mbox{\boldmath $\nabla$} m^*)^2
\nonumber \\[3mm]
& & \mbox{}
- \lf. 2 \frac{m^*}{k_{{\rm{F}}q}}
\Delta m^* \ri] ,
\label{eqB2} \eeqa
with $k_{{\rm F}q}^2 = (\mu_q - V_q)^2 - {m^*}^2$ and
$V_q= g\dr{v} V + g_\rho \tau_3 R/2$ (see Appendix A for notation).

Since in the semi-infinite geometry $\mbox{\boldmath $\nabla$}
\rightarrow d/dz$ and $\Delta \rightarrow d^2/dz^2 + \kappa d/dz$, one
easily finds
%
%Equation B3
\beqa
\frac{\partial \rho_{q,\,{\rm nr}}\ur{WK}}{\partial \kappa}
 & = & - \frac{1}{24 \pi^2}
\lf(\frac{2m}{\hbar^2}\ri)^{1/2} \lf[
5\frac{f^\prime_{q}}{f_{q}}\lf(\frac{\mu_q -
V_{q}}{f_{q}} \ri)^{1/2} + \frac{V^\prime_{q}}{f_{q}} \lf(
\frac{\mu_q -V_{q}}{f_{q}} \ri)^{-1/2} \ri]
\label{eqB3} \\[3mm]
%
%Equation B3b
\frac{\partial \rho_{q,\,{\rm rel}}\ur{WK}}{\partial \kappa}
 & = & - \frac{1}{12 \pi^2}
 \lf[ \lf( \frac{\ep_{{\rm{F}}q}}{k_{{\rm{F}}q}}
+ 2 \ln \frac{k_{{\rm{F}}q}+\ep_{{\rm{F}}q}}{m^*} \ri) V_q^\prime
+ \frac{m^*}{k_{{\rm{F}}q}} \, m^{*\prime} \ri] ,
\label{eqB3b} \eeqa
where the primes denote a derivative with respect to $z$.

Following standard techniques to pass from the WK expressions to the
ETF or RETF functionals (see e.g.\ Refs.\
\cite{Gra79,Rin80,Bra85,Cen93b,Spe93}), it is possible to eliminate
algebraically the derivatives of the potential $V_q$ in favour of the
derivatives of the density $\rho_q$ and the effective mass. This way,
from Eqs.\ \req{eqB3} and \req{eqB3b} one gets the corresponding ETF
and RETF contributions:
%
%Equation B5
\beqa
\lf( \frac{\partial \rho_{q}}{\partial \kappa}\ri)\dr{ETF} & = &
\frac{1}{12 \pi^2} (3 \pi^2 \rho_q)^{1/3}
 \lf( \frac{1}{3} \frac{\rho^\prime_q}{\rho_q}
 - 2 \frac{f^\prime_q}{f_q} \ri)
\label{eqB5} \\[3mm]
%
%Equation B5b
\lf( \frac{\partial \rho_{q}}{\partial \kappa}\ri)\dr{RETF} & = &
\frac{1}{12 \pi^2}
\lf[ \frac{\pi^2}{k_{{\rm{F}}q}^2} \lf( 1
+ 2 \frac{k_{{\rm{F}}q}}{\ep_{{\rm{F}}q}}\ln \frac{k_{{\rm{F}}q} +
\ep_{{\rm{F}}q}}
{m^*}\ri) \rho_q^\prime \ri.
\nonumber \\[3mm]
& & \mbox{}
+ \lf. 2 \lf( \frac{m^*}{\ep_{{\rm{F}}q}}
 \ln \frac{k_{{\rm{F}}q} + \ep_{{\rm{F}}q}}{m^*} \ri)
 m^{* \prime} \ri] .
\label{eqB5b} \eeqa
Finally, replacing these results into Eq.\ \req{eq46} one is able to
calculate $E\dr{c,e}\ur{dyn}$ for Skyrme forces and the \sw model.

\pagebreak
% >>>>>>>>>>>>>>>>>>>>>>>>>>>>>>>>>>>>>>>>>>>>>>>>>>>>>>>>>>>>>>>>>>>>
% REFERENCES.
%

%
\pagebreak
% >>>>>>>>>>>>>>>>>>>>>>>>>>>>>>>>>>>>>>>>>>>>>>>>>>>>>>>>>>>>>>>>>>>>
% TABLE CAPTIONS.
%
\section*{Table captions}
\begin{description}
\item[Table 1.]
Properties of infinite nuclear matter for the forces used in this
work. The last two rows show the values $\bar\delta\dr{nd}$ and
$\bar\delta\dr{pd}$ of the relative neutron excess at the neutron
and proton drip points.
\item[Table 2.]
Properties of symmetric semi-infinite nuclear matter (surface and
curvature energy coefficients $E\dr{s}$ and $E\dr{c}$, and surface
thickness $t$), plus the surface-stiffness coefficient $Q$. The labels
(1) and (2) for $Q$ stand for Eqs.\ \req{eq48} and \req{eq52}.
Besides, we present results from previous works: Ref.\ \cite{Kol85}
(KPLT) and Ref.\ \cite{Von94b} (VPSW). The calculations of VPSW were
performed in the Thomas--Fermi approximation.
\item[Table 3.]
Change of the surface properties of symmetric matter, and of
$\bar\delta\dr{nd}$ and $Q$, with the properties of the relativistic
interaction. The first row gives the results for the non-linear set
NLM\@. This set is defined by $a\dr{v}= -16$ MeV, $\rho\dr{nm}= 0.16$
fm$^{-3}$, $K= 200$ MeV, $m_\infty^*/m= 0.70$, $J= 30$ MeV and
$m\dr{s}= 500$ MeV\@. (The nucleon and $\omega$- and $\rho$-meson
masses are $m= 939$ MeV, $m\dr{v}= 783$ MeV and $m_\rho= 763$ MeV\@.)
The next rows correspond to sets of parameters that differ from NLM
only by the property listed in the first column.
\item[Table 4.]
Same as Table 3 for the surface properties of asymmetric matter at
$\delta_0= 0.212$ (below neutron drip) and $\delta_0= 0.6$ (above
neutron drip). Units are MeV for the energies and fm for $t$ and
$\Theta$.
\item[Table 5.]
Energy per nucleon of uncharged large nuclei with mass number $A$ and
overall neutron excess $I= 0.2$, for the Skyrme force SkM*. From a
semiclassical calculation of finite nuclei we obtain the asymmetry
$\delta_0$ at the center of the nucleus, Eq.\ \req{eq4.19}, and the
energy $E\dr{FN}(I)$. $E\dr{MF,e}(\delta_0)$ and
$E\dr{MF,\mu}(\delta_0)$ are the results of the mass formulae
\req{eq4.8} and \req{eq4.9}, including a term $-(2E\dr{s,0}^2/K)$.
Finally, $E\dr{LDM}(I)$ is the prediction of the LDM mass formula
\req{eq4.13}.
\item[Table 6.]
Same as Table 5 for the relativistic set NL2.
\end{description}

\pagebreak
% >>>>>>>>>>>>>>>>>>>>>>>>>>>>>>>>>>>>>>>>>>>>>>>>>>>>>>>>>>>>>>>>>>>>
% THE TABLES.
%
\section*{Table 1}
\vspace{2cm}
\begin{center}  \small
\begin{tabular}{lcccccccccc}
\hline
  & & SkM* & & SIII & & NL1 & & NL2 & & NL-SH  \\
\hline
$a\dr{v}$ (MeV) & & $-$15.77 & & $-$15.85 & & $-$16.42 & & $-$17.02
& & $-$16.35 \\
$\rho\dr{nm}$ (fm$^{-3}$) & & 0.160 & & 0.145 & & 0.152 & &  0.146
& & 0.146 \\
$K$ (MeV)       & & 216.6 & & 355.4 & & 211.1 & &  399.2 & &
355.3  \\
$m^{*}_{\infty}/m$ & & 0.789 & & 0.763 & & 0.573 & &  0.670 & &
0.598 \\
$J$ (MeV)   & & 30.03 & & 28.16 & & 43.46 & &  45.12 & & 36.12 \\
$L$ (MeV)   & & 45.8  & &  9.9  & & 140.2 & & 133.4  & & 113.7 \\
\hline
$\bar{\delta}\dr{nd}$  & & 0.322 & & 0.337 & & 0.240 & &  0.222 & &
0.284 \\
$\bar{\delta}\dr{pd}$ & & 0.820 & &  0.970 & & 0.637 & &  0.749 & &
0.714 \\
\hline
\end{tabular}
\end{center}
\pagebreak
\section*{Table 2}
\vspace{1cm}
\begin{center}  \small
\begin{tabular}{lcccccccccc}
\hline
    & & SkM* & & SIII & & NL1 & & NL2 & & NL-SH  \\
\hline
$E\dr{s}$ (MeV) \\
  this work   & & 16.00 & & 16.47 & & 17.38 & & 19.65 & & 17.22 \\
  KPLT (1/36) & & 16.05 & & 16.55 & &    & &    & &    \\
KPLT (1/18)   & & 17.96 & & 18.79 & &    & &    & &    \\
  VPSW (TF)   & &       & &       & & 19.78    & &    & & 20.07 \\
\hline
$E\dr{c}$ (MeV) \\
  this work & & 10.53 & & 7.33 & & 12.59 & & 9.10 & & 8.40 \\
KPLT (1/36) & & 10.74 & &  7.50 & &    & &    & &     \\
KPLT (1/18) & & 13.87 & & 10.10 & &    & &    & &     \\
\hline
$t$               (fm) \\
  this work    & &  2.23 & &  1.72 & &  2.11 & &  1.50 & &  1.50 \\
KPLT (1/36)    & &  2.26 & &  1.75 & &    & &    & &     \\
KPLT (1/18)    & &  2.45 & &  1.93 & &    & &    & &     \\
VPSW (TF)      & &       & &       & &  2.90  & &   & & 2.09 \\
\hline
$Q$ (MeV) \\
this work (1)  & & 39.6  & & 63.8  & & 29.8 & & 41.9 & & 34.7 \\
this work (2)  & & 38.4  & & 63.3  & & 29.0 & & 41.6 & & 34.2 \\
KPLT (1/36)    & & 38  & & 61      & &      & &    & &      \\
KPLT (1/18)    & & 34  & & 54      & &      & &    & &      \\
VPSW (TF)      & &     & &         & & 24.4 & &    & & 27.6 \\
\hline
\end{tabular}
\end{center}
\pagebreak
\section*{Table 3}
\vspace{1cm}
\begin{center}  \small
\begin{tabular}{lccccc}
\hline
  & &   \mcl{3}{c}{$\delta_0 = 0$} & \\
      \cline{3-5}
   & $\bar\delta\dr{nd}$ & $E\dr{s}$ (MeV) &
      $E\dr{c}$ (MeV)    & $t$ (fm)        &
      $Q$ (MeV) \\
\hline
 Set NLM             & 0.385  & 14.81 &  9.53 & 1.68 & 28.8 \\
$a\dr{v} = -17$ MeV  & 0.439  & 14.67 &  9.26 & 1.59 & 30.4 \\
$\rho\dr{nm} = 0.145$ fm$^{-3}$
                     & 0.387  & 14.29 &  8.88 & 1.68 & 29.4 \\
$K = 300$ MeV        & 0.355  & 17.56 &  9.61 & 1.61 & 30.0 \\
$m^*_\infty/m = 0.55$
                     & 0.446  & 16.54 & 12.01 & 2.03 & 22.2 \\
$J = 40$ MeV         & 0.252  & 14.81 &  9.53 & 1.68 & 35.0 \\
$m\dr{s} = 550$ MeV  & 0.385  & 11.82 &  6.05 & 1.18 & 38.4 \\
\hline
\end{tabular}
\end{center}
\vspace*{1cm}
\section*{Table 4}
\vspace{1cm}
\begin{center}  \small
\begin{tabular}{lccccccccccccc}
\hline
   &    \mcl{6}{c}{$\delta_0= 0.212$}
 & &    \mcl{6}{c}{$\delta_0= 0.6$} \\
      \cline{2-7} \cline{9-14}
   &
$E\dr{s,e}$ & $E\dr{c,e}$ & $E\dr{s,\mu}$ & $E\dr{c,\mu}$ & $t$ &
$\Theta$
   & &
$E\dr{s,e}$ & $E\dr{c,e}$ & $E\dr{s,\mu}$ & $E\dr{c,\mu}$ & $t$ &
$\Theta$
 \\
\hline
 Set NLM &
  18.7 & 11.6 & 12.4 &  8.1 & 2.06 & 0.43 & &
  17.0 & 18.8 &  1.65 &  1.00 & 4.63 & 0.70 \\
$a\dr{v} = -17$ MeV &
  18.4 & 11.2 & 12.4 &  8.0 & 1.94 & 0.40 & &
  17.0 & 18.5 &  1.64 &  0.95 & 4.45 & 0.70 \\
$\rho\dr{nm} = 0.145$ fm$^{-3}$ &
  18.1 & 10.9 & 11.9 &  7.5 & 2.06 & 0.43 & &
  16.0 & 17.9 &  1.58 &  0.94 & 4.63 & 0.68 \\
$K = 300$ MeV &
  21.2 & 11.6 & 15.0 &  8.1 & 1.94 & 0.41 & &
  22.6 & 24.7 &  2.41 &  1.01 & 4.17 & 0.82 \\
$m^*_\infty/m = 0.55$ &
  21.7 & 15.1 & 13.6 & 10.6 & 2.58 & 0.56 & &
  19.3 & 21.4 &  1.58 &  1.23 & 6.20 & 0.91 \\
$J = 40$ MeV &
  21.0 & 14.4 & 11.2 &  6.8 & 2.26 & 0.51 & &
  15.1 & 19.2 &  1.40 &  0.99 & 4.89 & 0.48 \\
$m\dr{s} = 550$ MeV &
  14.9 &  8.4 & 10.0 &  5.4 & 1.51 & 0.33 & &
  13.8 & 15.2 &  1.36 &  0.73 & 3.69 & 0.56 \\
\hline
\end{tabular}
\end{center}
\pagebreak
\section*{Table 5}
\vspace{1cm}
\begin{center}  \small
\begin{tabular}{rcccccc}
\hline
    \mcl{7}{c}{SkM* \ \ ($I=0.2$)} \\
\hline
\mcl{1}{c}{$A$} & & $\delta_0$ & $E\dr{FN}(I)$ &
 $E\dr{MF,e}(\delta_0)$ &
 $E\dr{MF,\mu}(\delta_0)$ & $E\dr{LDM}(I)$  \\
\hline
   250 & & 0.152 & $-$12.129 & $-$12.124 & $-$12.171  &  $-$12.111 \\
   500 & & 0.160 & $-$12.664 & $-$12.667 & $-$12.652  &  $-$12.561 \\
  1000 & & 0.168 & $-$13.080 & $-$13.085 & $-$13.074  &  $-$13.068 \\
  5000 & & 0.180 & $-$13.723 & $-$13.728 & $-$13.722  &  $-$13.711 \\
 10000 & & 0.184 & $-$13.905 & $-$13.909 & $-$13.904  &  $-$13.892 \\
 20000 & & 0.187 & $-$14.047 & $-$14.049 & $-$14.047  &  $-$14.035 \\
\hline
\end{tabular}
\end{center}
\vspace*{2cm}
\section*{Table 6}
\vspace{1cm}
\begin{center}  \small
\begin{tabular}{rcccccc}
\hline
   \mcl{7}{c}{NL2 \ \ ($I=0.2$)}  \\
\hline
\mcl{1}{c}{$A$} & & $\delta_0$ & $E\dr{FN}(I)$ &
 $E\dr{MF,e}(\delta_0)$ &
 $E\dr{MF,\mu}(\delta_0)$ & $E\dr{LDM}(I)$  \\
\hline
   250 & & 0.137 & $-$12.447 & $-$12.456 & $-$12.422  &  $-$12.530 \\
   500 & & 0.147 & $-$13.064 & $-$13.079 & $-$13.052  &  $-$13.111 \\
  1000 & & 0.156 & $-$13.540 & $-$13.554 & $-$13.534  &  $-$13.564 \\
  5000 & & 0.172 & $-$14.276 & $-$14.285 & $-$14.275  &  $-$14.267 \\
 10000 & & 0.177 & $-$14.483 & $-$14.493 & $-$14.482  &  $-$14.466 \\
 20000 & & 0.181 & $-$14.646 & $-$14.659 & $-$14.646  &  $-$14.624 \\
\hline
\end{tabular}
\end{center}

\pagebreak
% >>>>>>>>>>>>>>>>>>>>>>>>>>>>>>>>>>>>>>>>>>>>>>>>>>>>>>>>>>>>>>>>>>>>
% FIGURE CAPTIONS.
%
\section*{Figure captions}
\begin{description}
\item[Figure 1.]
Neutron and proton chemical potentials of bulk matter ($\mu\dr{n}$ and
$\mu\dr{p}$) as a function of the relative neutron excess $\delta$,
for the relativistic interaction NL1. The vertical slashes indicate
the neutron and proton drip points ($\bar\delta\dr{nd}$ and
$\bar\delta\dr{pd}$) and the critical point ($\delta\dr{c}$) where the
densities of the two phases become equal. For $\delta < \delta\dr{c}$
it is $\delta = \delta_0$, the neutron excess of the nuclear phase.
When proton drip occurs ($\delta\dr{d} < 1$) the dashed lines allow
one to read the neutron excess $\delta\dr{d}$ of the drip phase that
is in equilibrium with the nuclear phase at $\delta_0$.
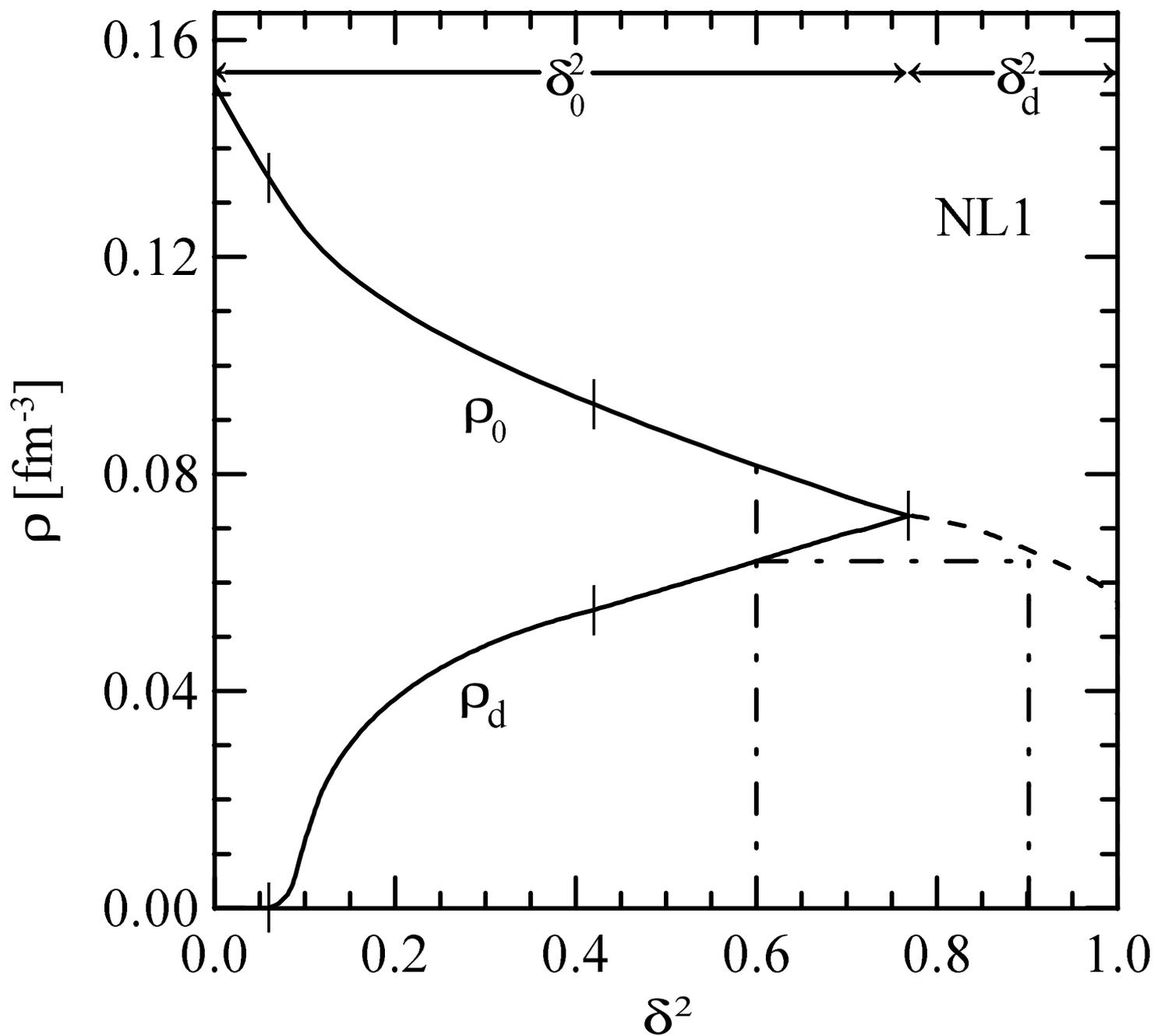
\item[Figure 2.]
Same as Fig.\ 1 for the densities of the nuclear and drip phases,
$\rho_0$ and $\rho\dr{d}$, against $\delta^2$. The superimposed
dot-dashed lines illustrate an example of coexistence between a
nuclear medium with $\delta_0^2= 0.6$ and density $\rho_0= 0.082$
fm$^{-3}$ and drip matter with $\delta\dr{d}^2= 0.9$ and density
$\rho\dr{d}= 0.064$ fm$^{-3}$.
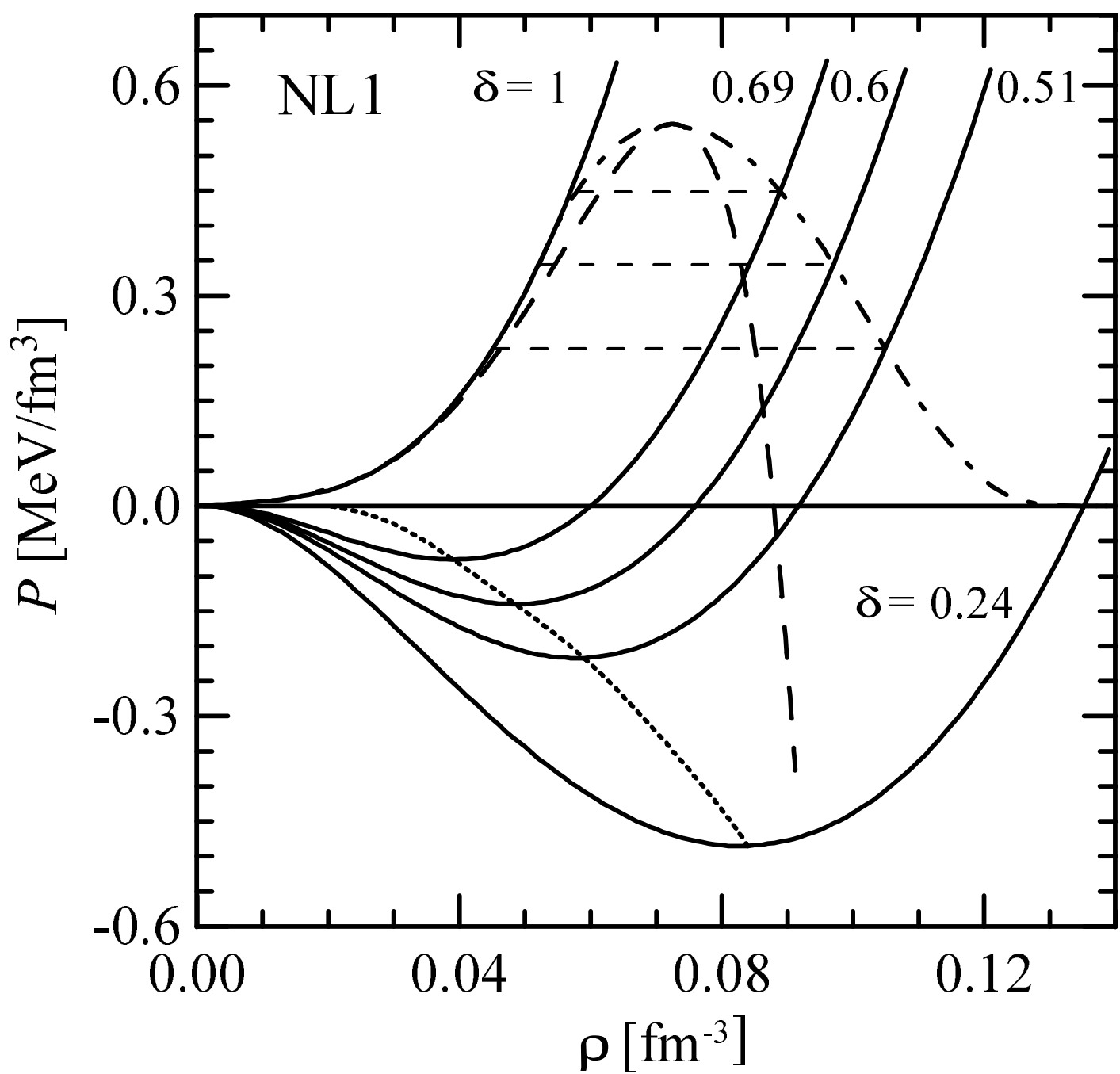
\item[Figure 3.]
Equation of state for the NL1 parametrization. The solid lines
represent the pressure $P$ as a function of the density $\rho$ for
several values of $\delta$ (from the neutron drip point to neutron
matter). The dot-dashed line defines the coexistence curve. The
long-dashed and short-dashed lines are the diffusive and mechanical
stability curves. Some samples of points on the coexistence curve that
can be in phase equilibrium have been joined by dashed lines.
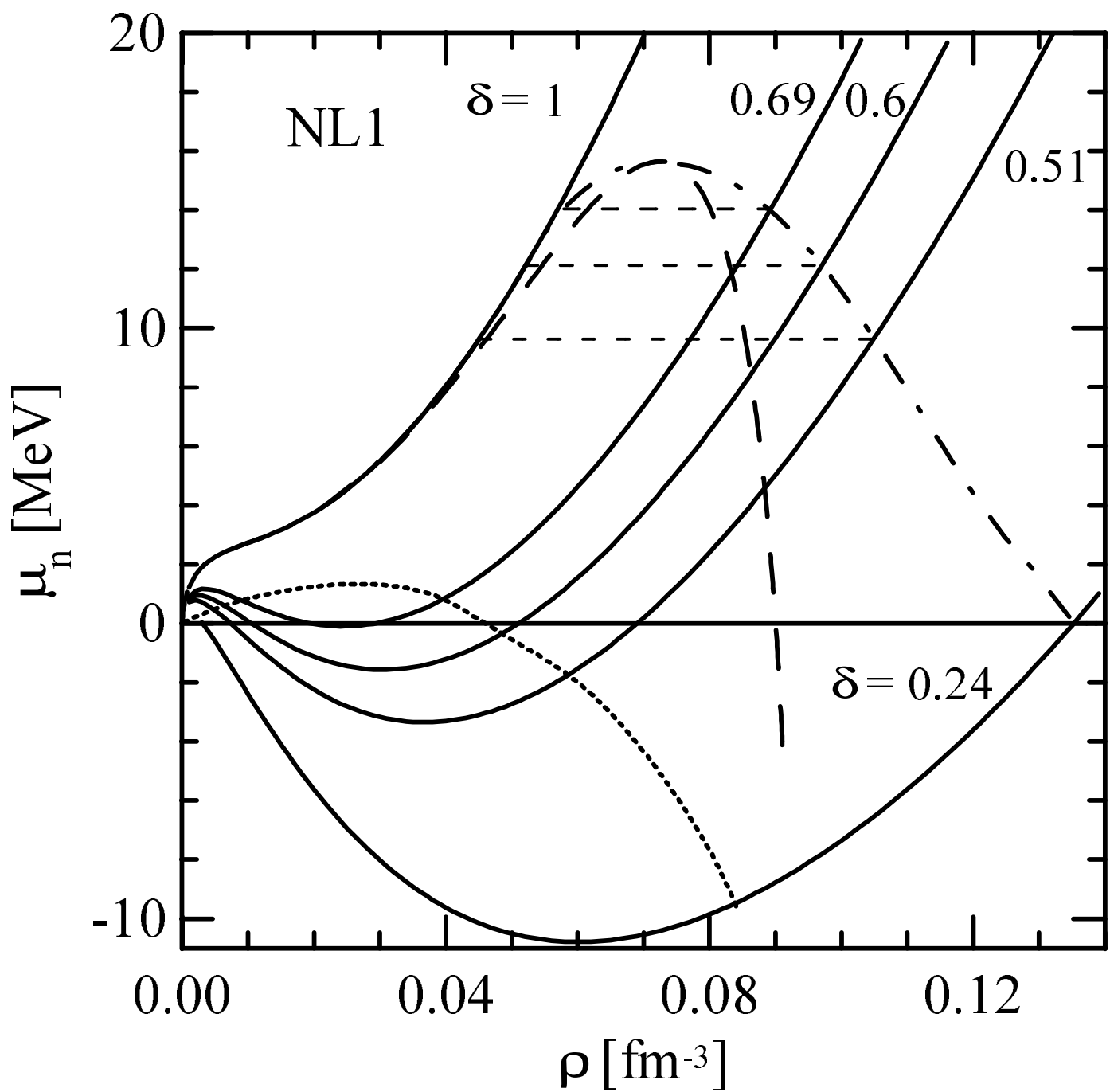
\item[Figure 4.]
Same as Fig.\ 3 for the neutron chemical potential $\mu\dr{n}$.
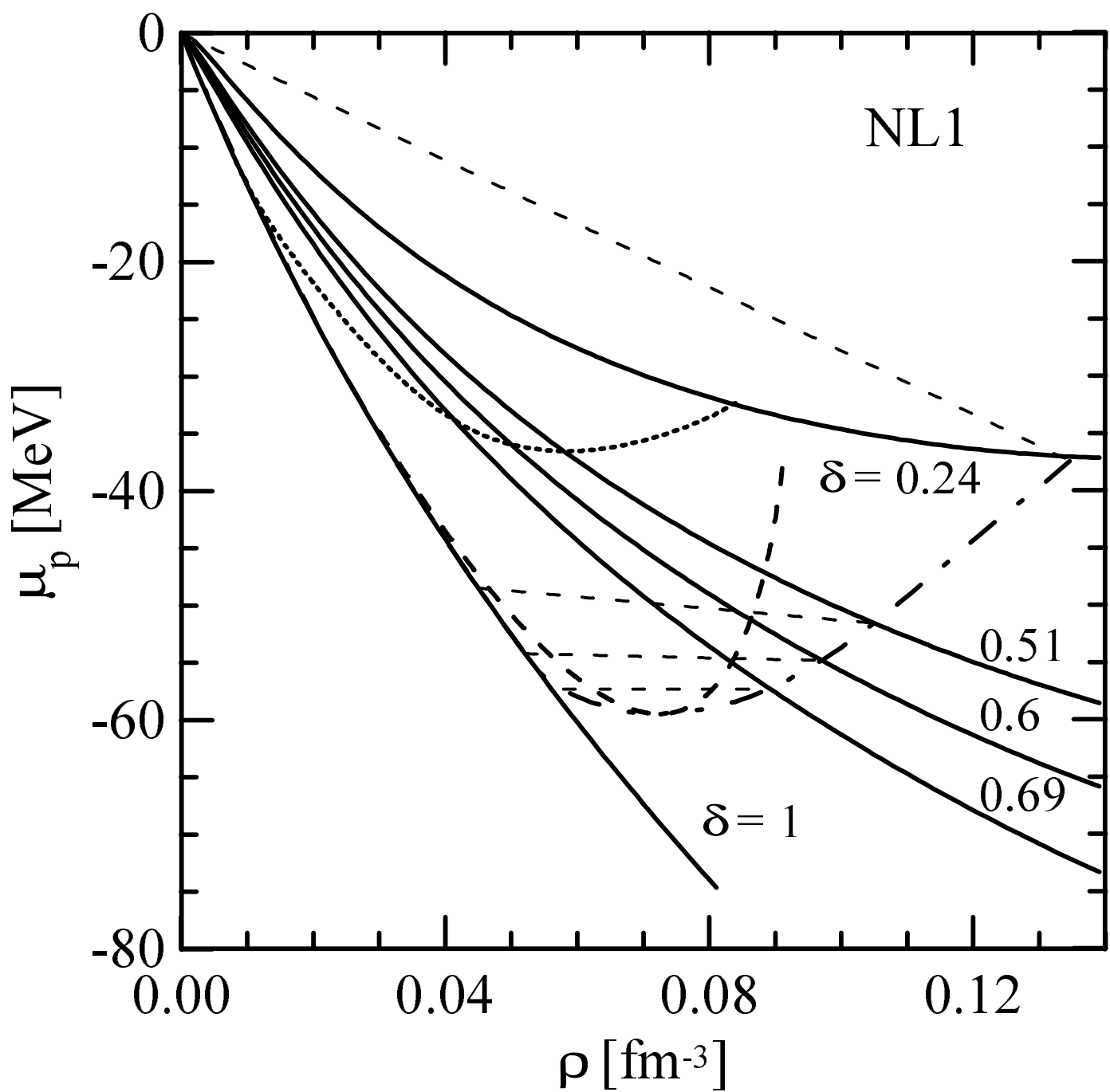
\item[Figure 5.]
Same as Fig.\ 3 for the proton chemical potential $\mu\dr{p}$.
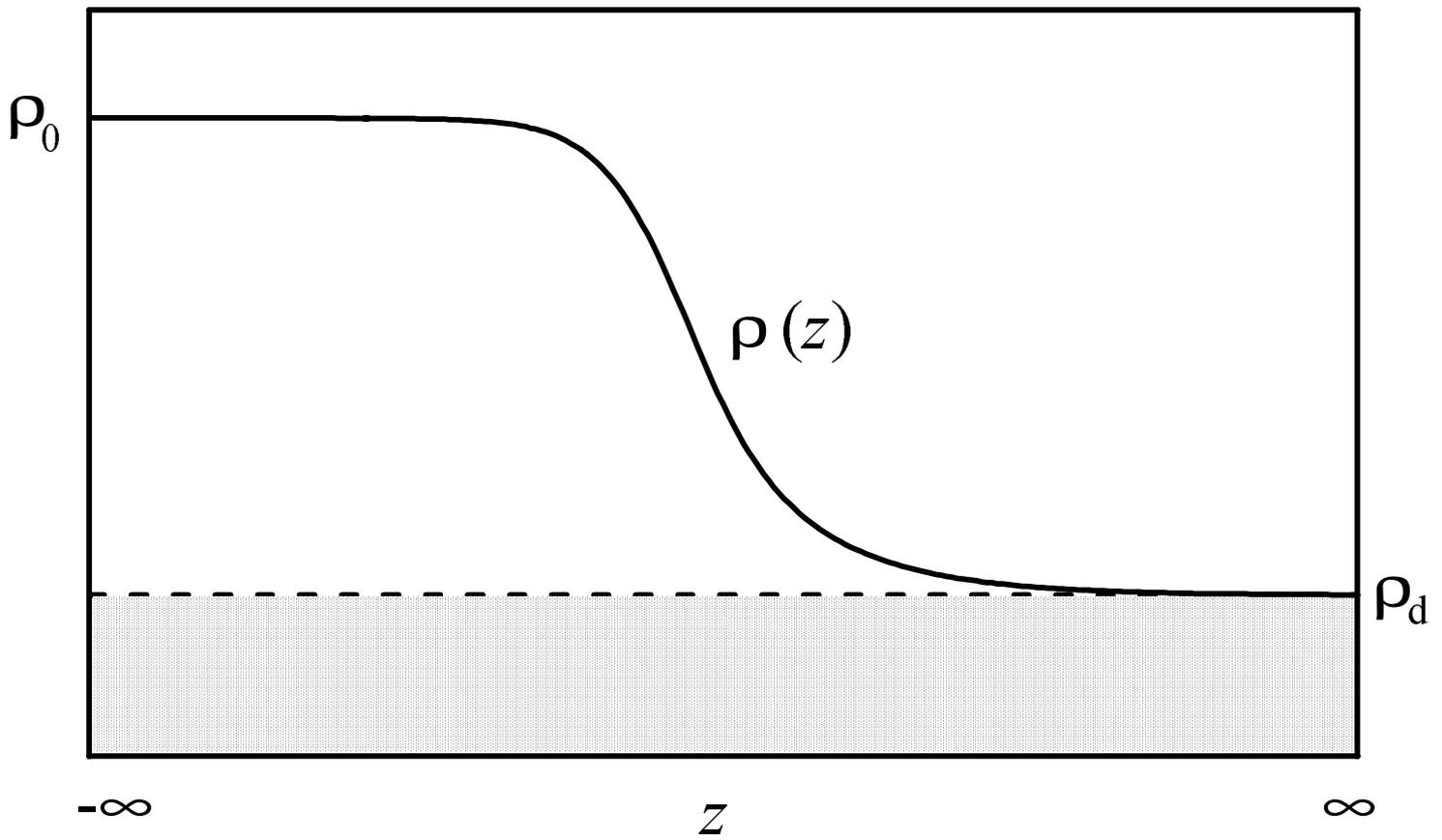
\item[Figure 6.]
Schematic representation of the density profile $\rho(z)$ of the
semi-infinite system with drip particles. The quantities $\rho_0$ and
$\rho\dr{d}$ are the asymptotic densities when $z \to -\infty$ and $z
\to \infty$, respectively.
\item[Figure 7.]
Neutron and proton local density profiles of semi-infinite matter for
the relativistic set NL1 and the Skyrme force SkM*. They are drawn for
several bulk asymmetries: $\delta_0= 0$ (symmetric system), $\delta_0=
0.24$ (below neutron drip), $\delta_0 = 0.4$ (above neutron drip), and
finally at a $\delta_0$ above proton drip. For the asymmetric cases
the lower curves are the proton densities. The vertical bars show the
surface thickness $t$, the distance where the density of the nucleus
drops from 90\% to 10\% of its central value.
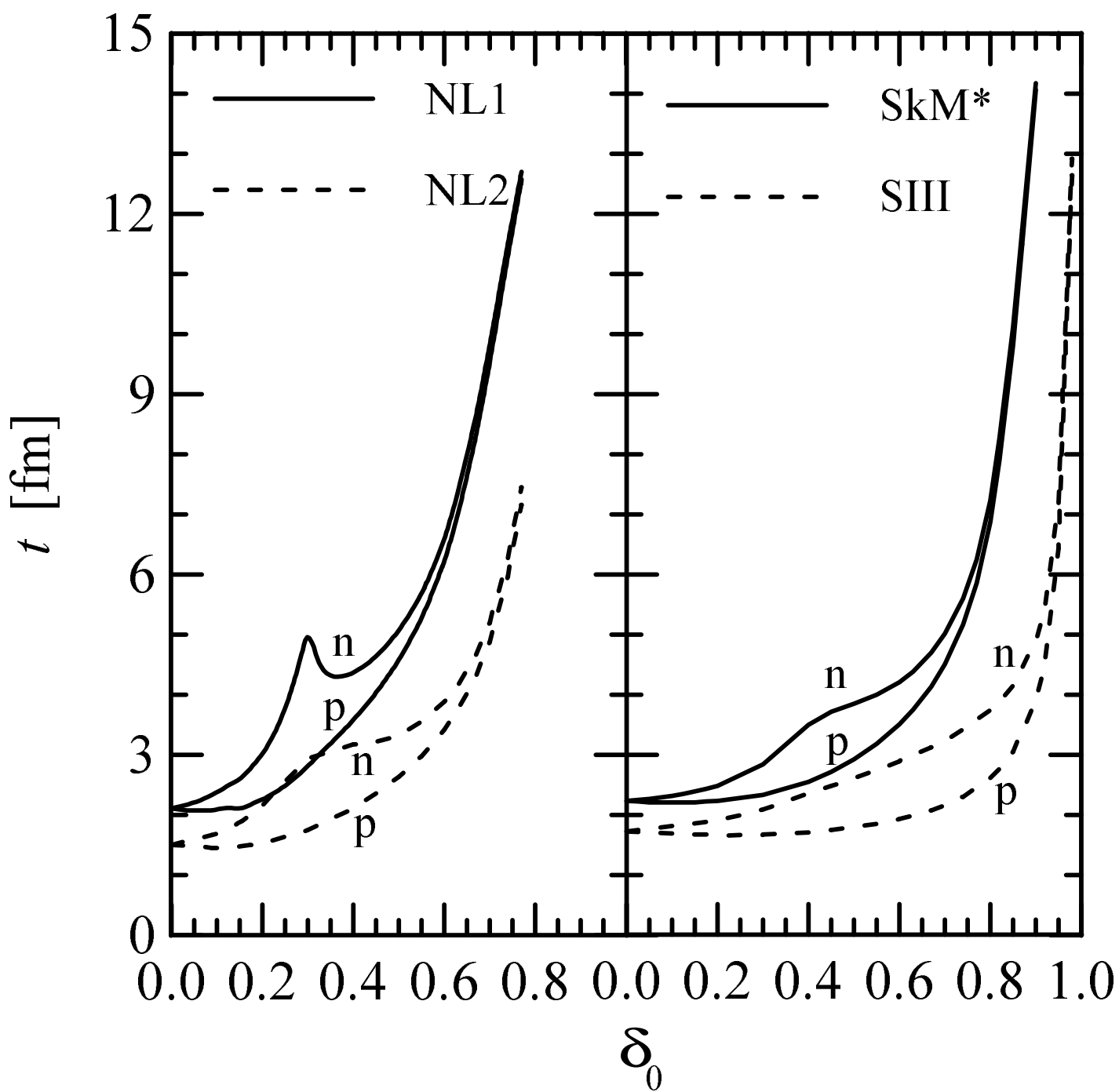
\item[Figure 8.]
Surface thickness $t$ of the neutron (n) and proton (p) density
distributions against the bulk neutron excess $\delta_0$ for the
relativistic sets NL1 and NL2 and for the Skyrme forces SkM* and
SIII\@.
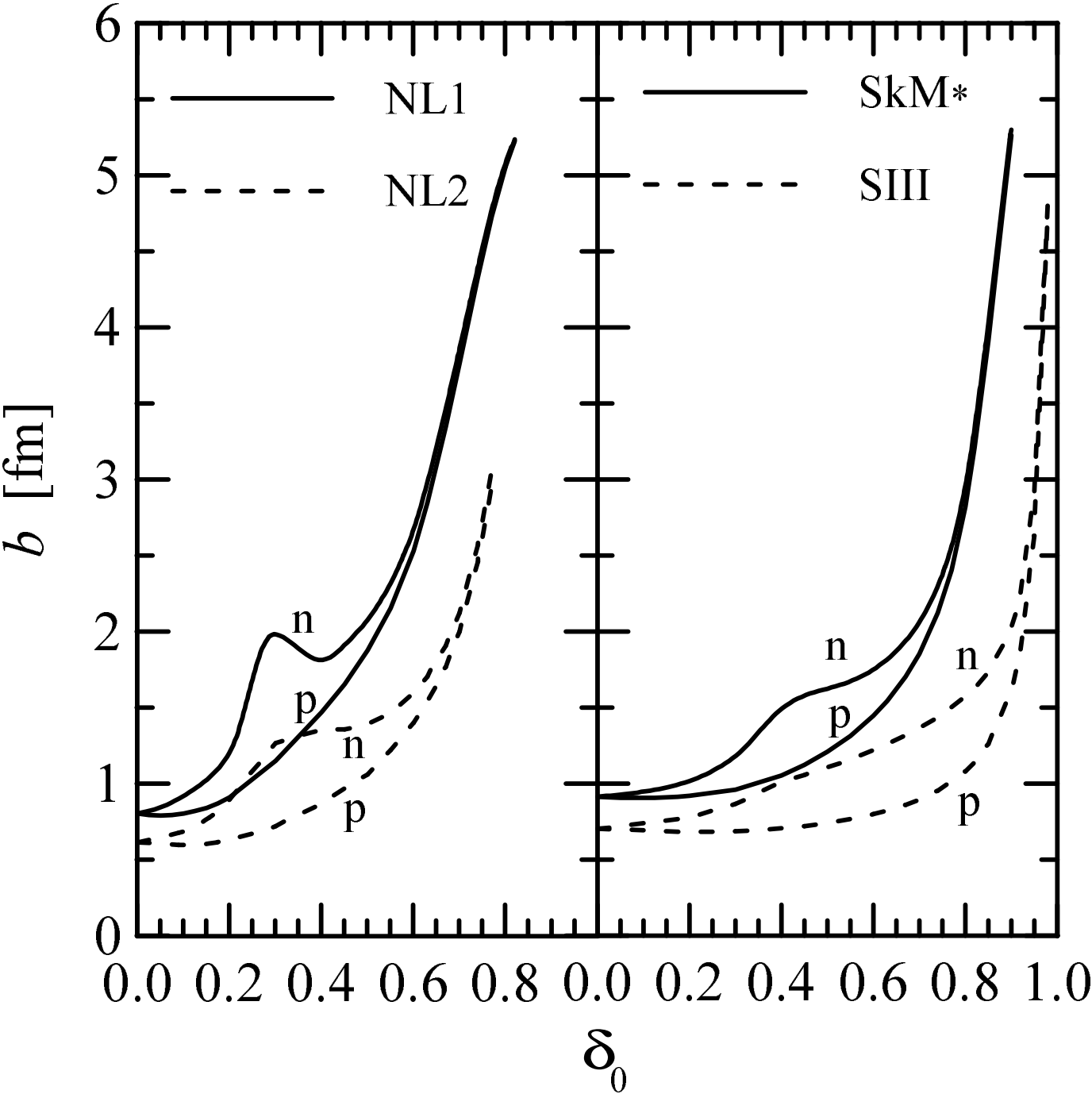
\item[Figure 9.]
Same as Fig.\ 8 for the surface width $b$ of the neutron and proton
density distributions.
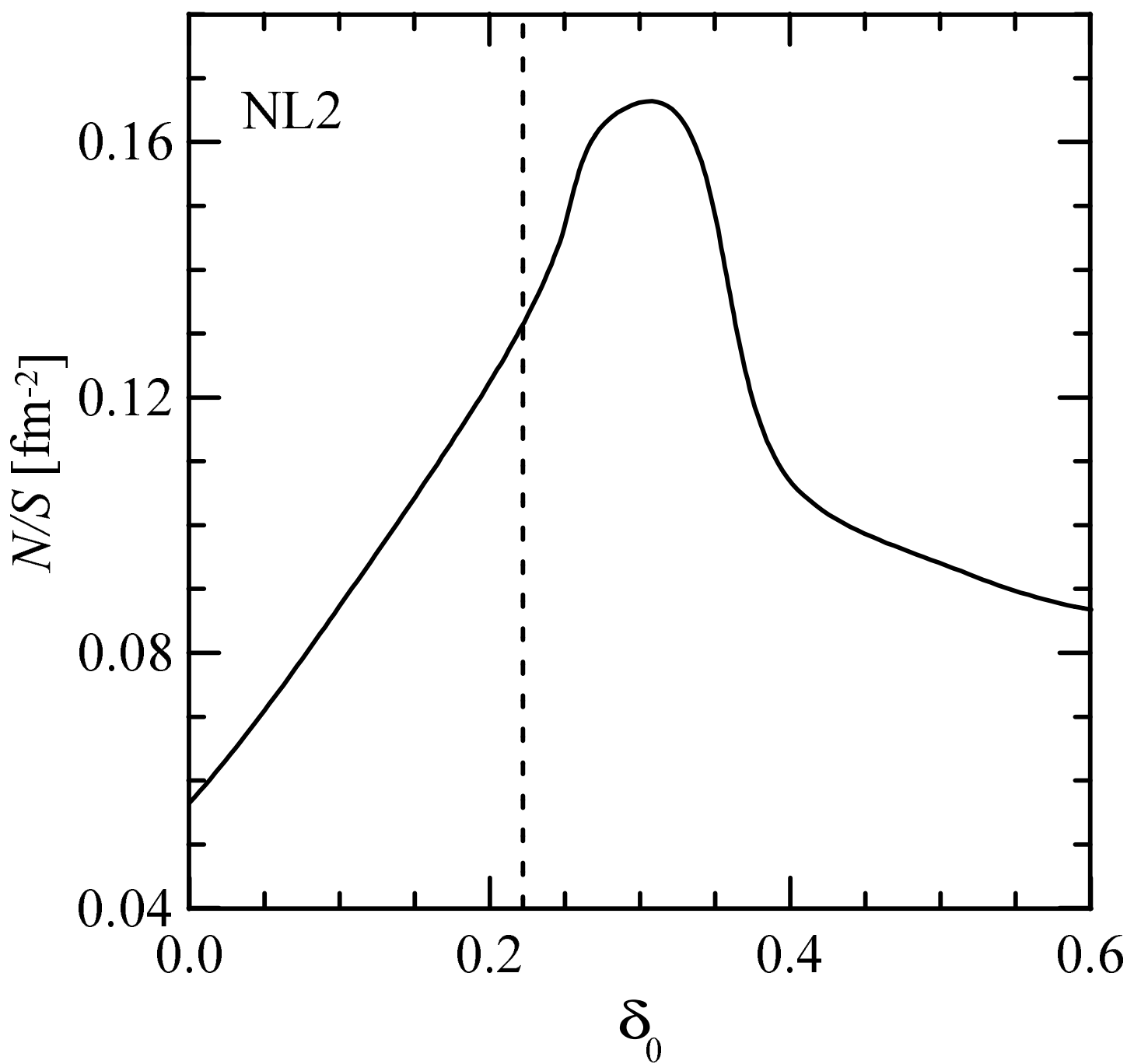
\item[Figure 10.]
Number of neutrons per unit area in the surface region (see text for
explanation) as a function of $\delta_0$ for the NL2 parametrization.
The vertical line indicates the neutron drip point.
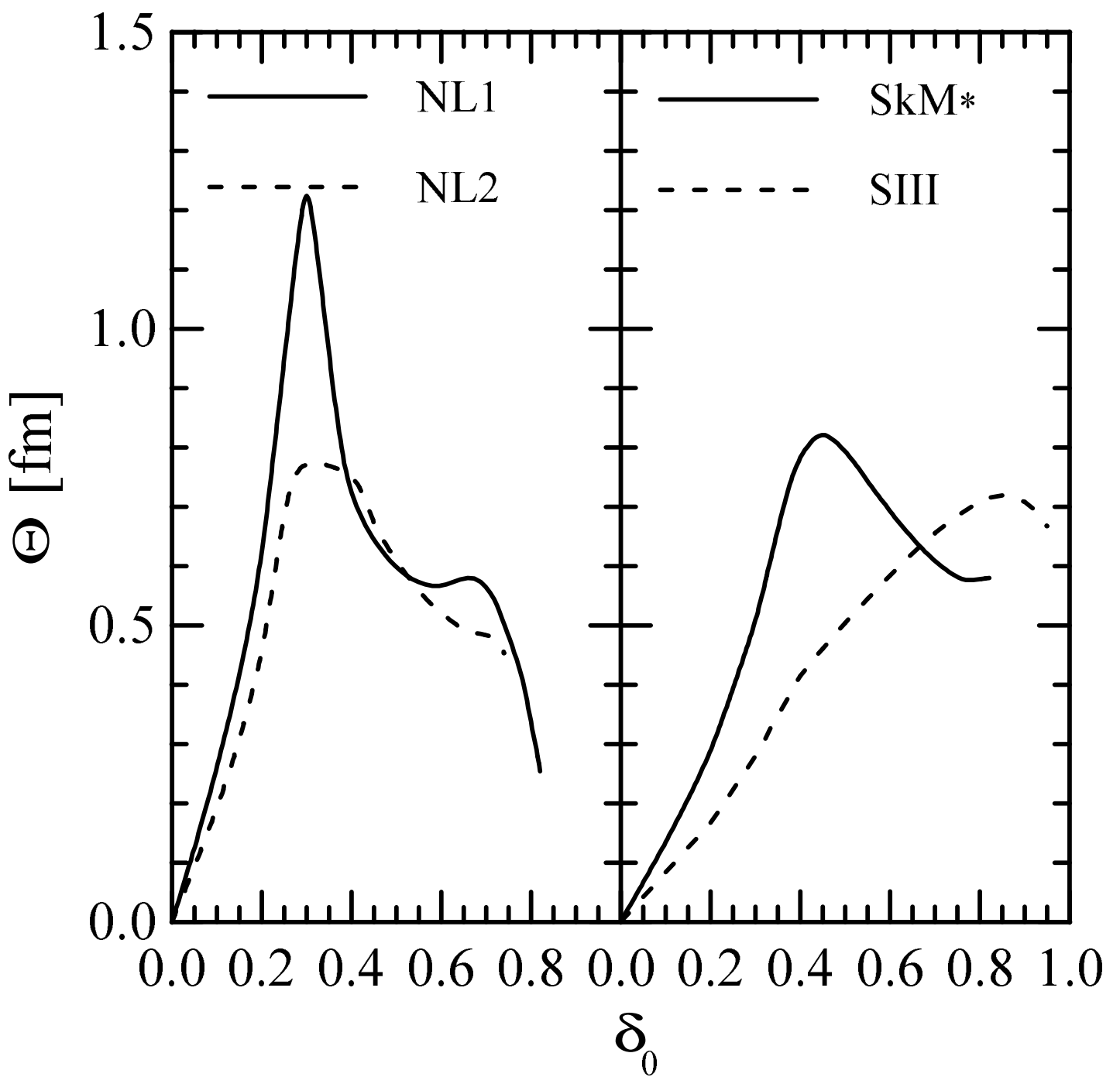
\item[Figure 11.]
Neutron skin thickness $\Theta$ versus $\delta_0$.
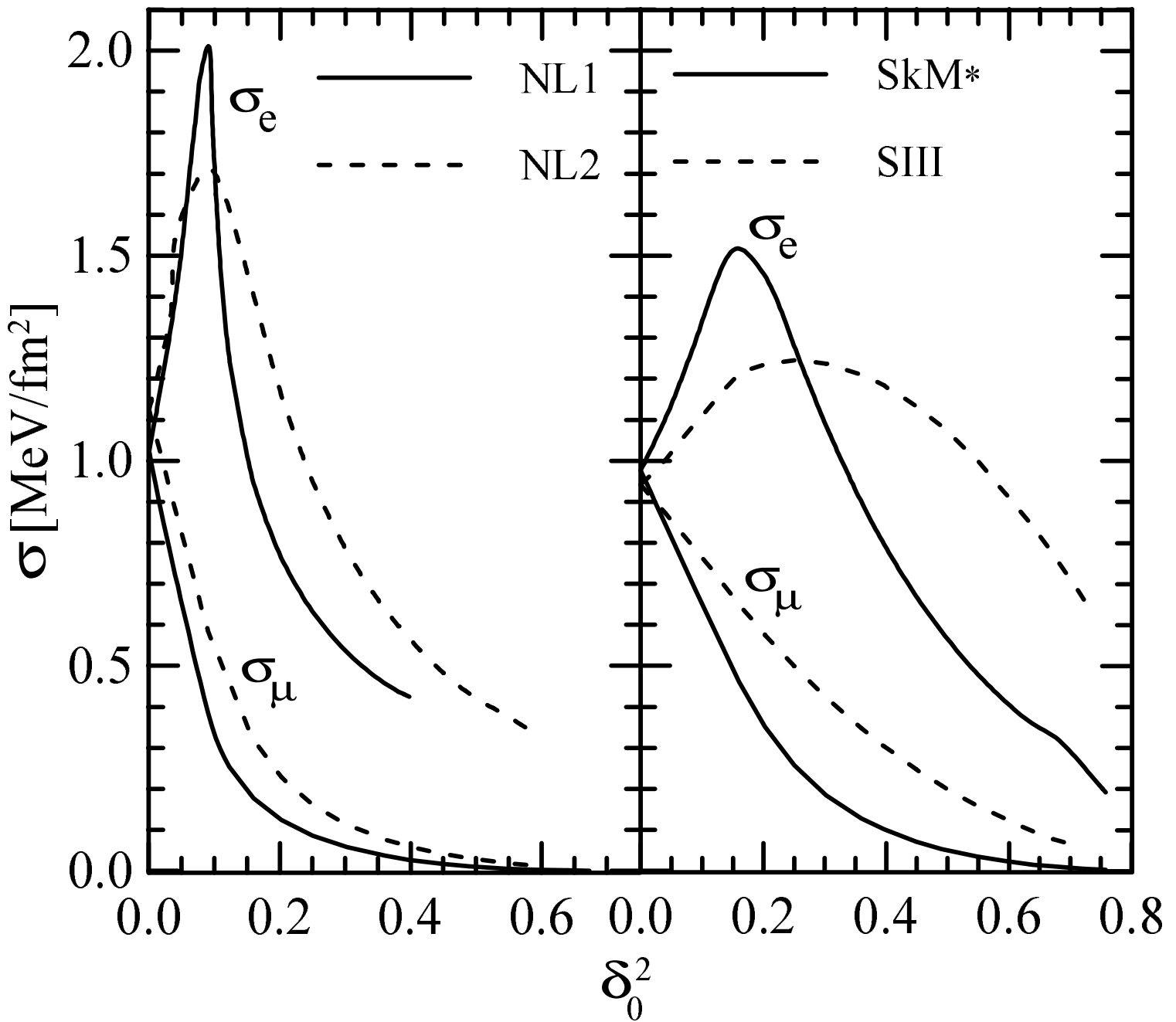
\item[Figure 12.]
Dependence of the surface tension upon $\delta_0^2$ for the two
definitions of the reference energy discussed in the text.
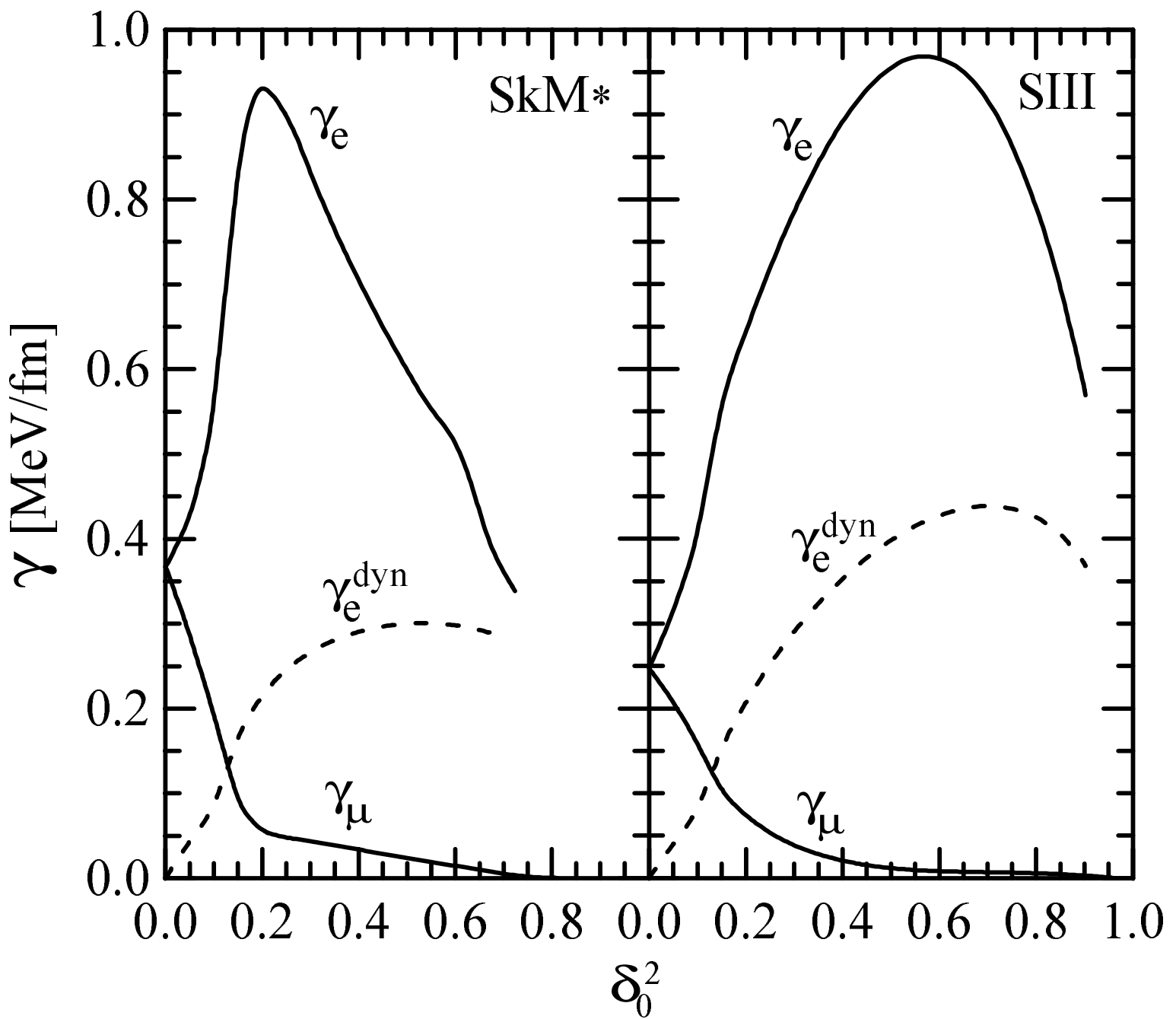
\item[Figure 13.]
Dependence of the curvature energies per unit length $\gamma\dr{e}$
and $\gamma_\mu$ upon $\delta_0^2$ for SkM* and SIII\@. The dashed
lines show the dynamical contribution $\gamma\dr{e}\ur{dyn}$.
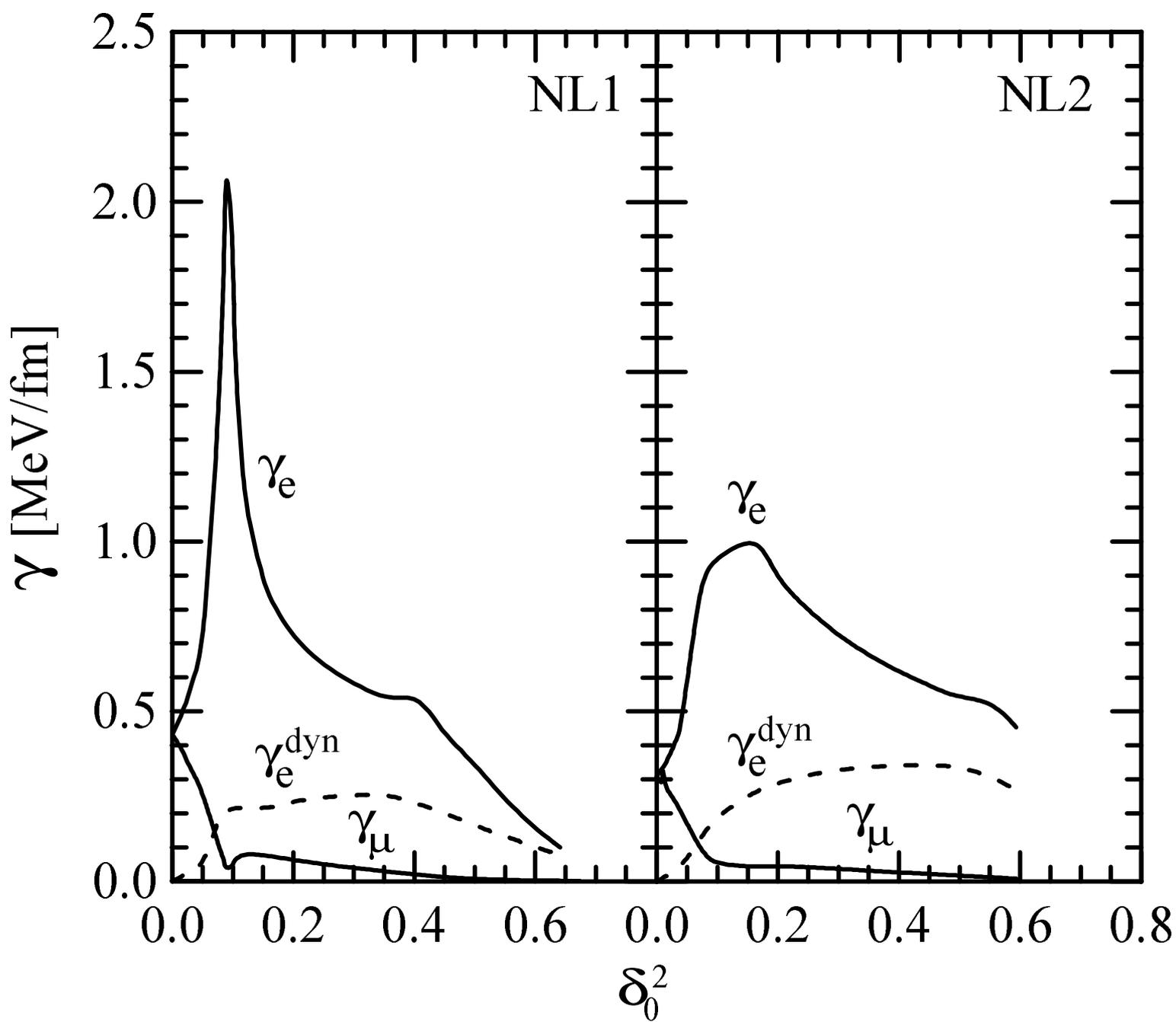
\item[Figure 14.]
Same as Fig.\ 13 for the relativistic sets NL1 and NL2.
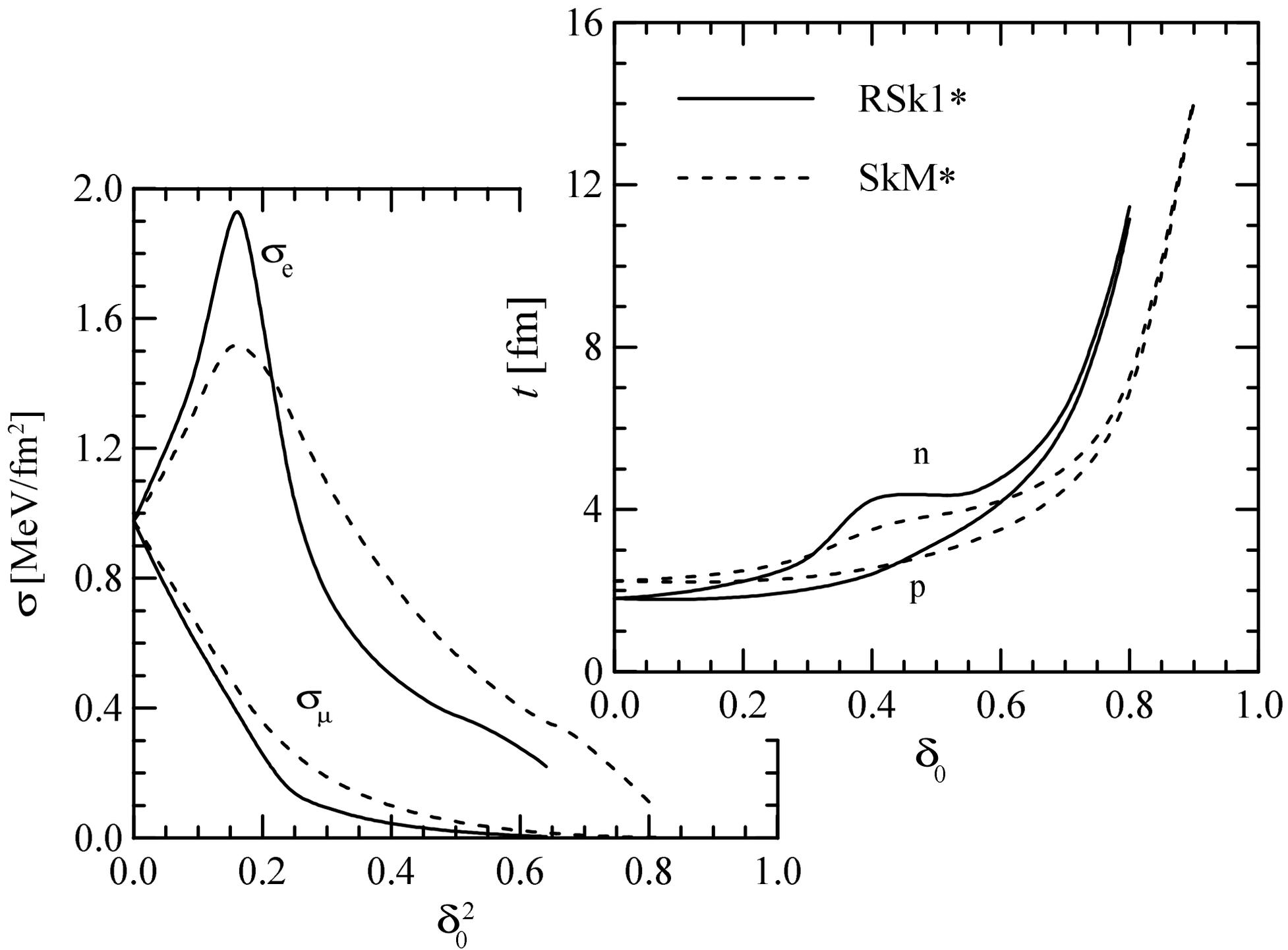
\item[Figure 15.]
Surface tension and surface thickness for SkM* and for the
relativistic set RSk1* adjusted to the nuclear matter properties and
surface energy at $\delta_0= 0$ of SkM*.
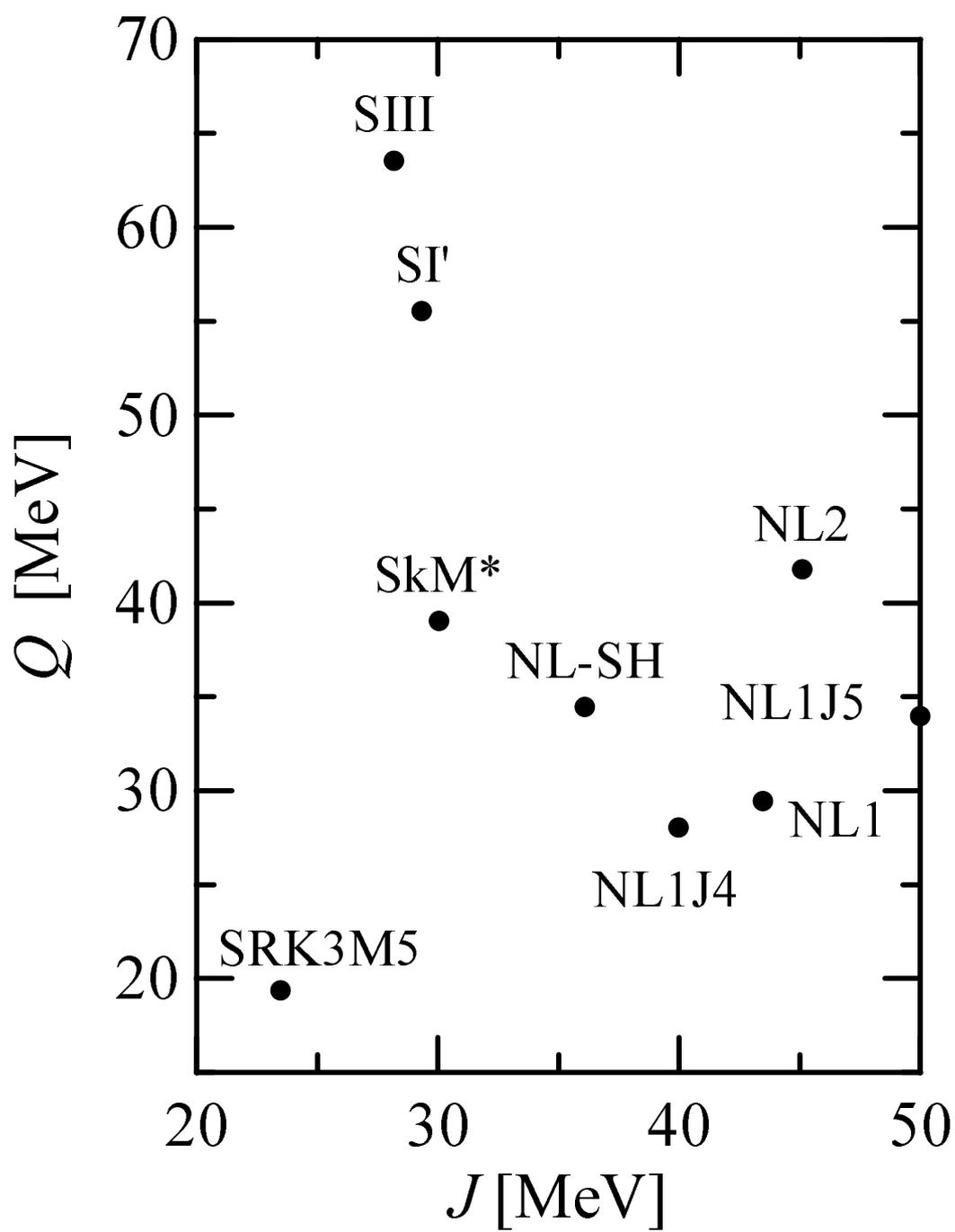
\item[Figure 16.]
Surface-stiffness coefficient $Q$ against the bulk symmetry energy $J$
for some relativistic and non-relativistic parametrizations.
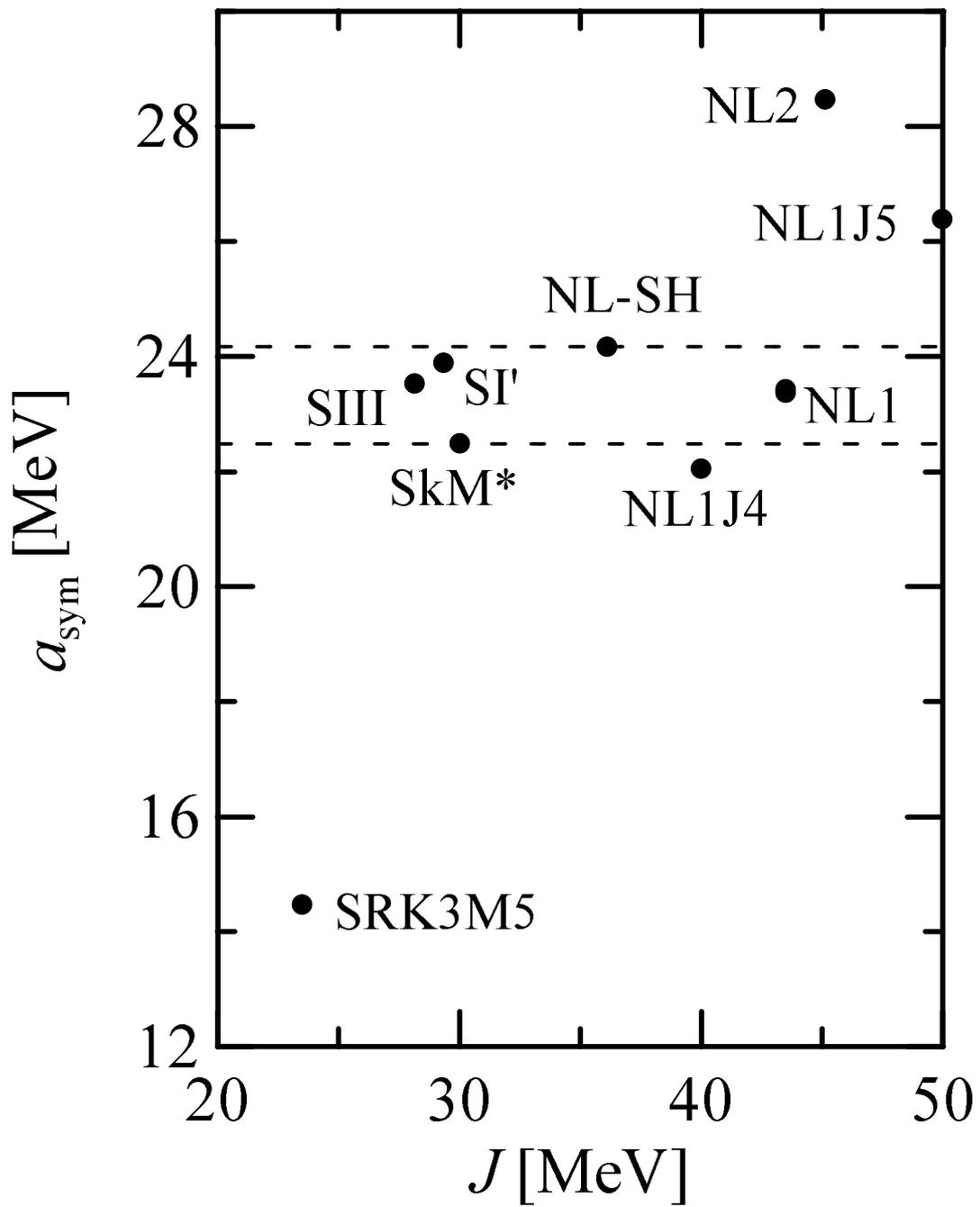
\item[Figure 17.]
Mass-formula symmetry coefficient $a\dr{sym}$ calculated for
$^{208}$Pb, Eq.\ \req{eq53}, against the bulk symmetry energy $J$ for
the forces of Fig.\ 16. The dashed lines roughly indicate the
empirical region.
\end{description}
%
% >>>>>>>>>>>>>>>>>>>>>>>>>>>>>>>>>>>>>>>>>>>>>>>>>>>>>>>>>>>>>>>>>>>>

\begin{thebibliography}{99}
%
%%%%%%%%%%%%%%%%
\parskip= -1.0mm
%%%%%%%%%%%%%%%%
%
\bibitem{Mye69} W.D. Myers and W.J. Swiatecki, \AP{55} (1969) 395;
                \AP{84} (1974) 186;
                W.D. Myers, Droplet model of atomic nuclei
                (Plenum, New York, 1977).
%
\bibitem{Gra79} B. Grammaticos and A. Voros, \AP{123} (1979) 359;
                \AP{129} (1980) 153.
%
\bibitem{Rin80} P. Ring and P. Schuck, The nuclear many-body problem
                (Springer, New York, 1980) ch.\ 13.
%
\bibitem{Bra85} M. Brack, C. Guet and H.-B. H{\aa}kansson,
                \PRp{123} (1985) 275.
%
\bibitem{Vau72} D. Vautherin and D.M. Brink, \PRC{5} (1972) 626.
%
\bibitem{Tre86} J. Treiner and H. Krivine, \AP{170} (1986) 406.
%
\bibitem{Sto88} W. Stocker, J. Bartel, J.R. Nix and A.J. Sierk,
                \NPA{489} (1988) 252.
%
\bibitem{Dur93} M. Durand, P. Schuck and X. Vi\~nas,
                \ZPA{346} (1993) 87.
%
\bibitem{Cen96} M. Centelles, X. Vi\~{n}as and P. Schuck,
                \PRC{53} (1996) 1018.
%
\bibitem{Mye85} W.D. Myers, W.J. Swiatecki and C.S. Wong, \NPA{436}
                (1985) 185.
%
\bibitem{Kol85} K. Kolehmainen, M. Prakash , J.M. Lattimer and J.
                Treiner, \NPA{439} (1985) 537.
%
\bibitem{Ser86} B.D. Serot and J.D. Walecka, \ANP{16} (1986) 1.
%
\bibitem{Cel86} L.S. Celenza and C.M. Shakin,
                Relativistic nuclear physics: theories of structure
                and scattering (World Scientific, Singapore, 1986).
%
\bibitem{Rei89} P.-G. Reinhard, \RPP{52} (1989) 439.
%
\bibitem{Ser92} B.D. Serot, \RPP{55} (1992) 1855.
%
\bibitem{Fri75} J.L. Friar and J.W. Negele, \ANP{8} (1975) 219.
%
\bibitem{Bog77} J. Boguta and A.R. Bodmer, \NPA{292} (1977) 413.
%
\bibitem{Hor81} C.J. Horowitz and B.D. Serot, \NPA{368} (1981) 503.
%
\bibitem{Bou84} A. Bouyssy, S. Marcos and Pham Van Thieu,
                \NPA{422} (1984) 541.
%
\bibitem{Rei86} P.-G. Reinhard, M. Rufa, J. Maruhn, W. Greiner and
                J. Friedrich, \ZPA{323} (1986) 13.
%
\bibitem{Gam90} Y.K. Gambhir, P. Ring and A. Thimet,
                \AP{198} (1990) 132.
%
\bibitem{Hof89}  D. Hofer and W. Stocker, \NPA{492} (1989) 637.
%
\bibitem{Von94a} D. Von-Eiff, W. Stocker and M.K. Weigel, \PRC{50}
                 (1994) 1436.
%
\bibitem{Bog77b} J. Boguta and J. Rafelski, \PLB{71} (1977) 22.
%
\bibitem{Sto91}  W. Stocker and M.M. Sharma, \ZPA{339} (1991) 147.
%
\bibitem{Sha91}  M.M. Sharma, S.A. Moszkowski and P. Ring,
                 \PRC{44} (1991) 2493.
%
\bibitem{Cen90} M. Centelles, X. Vi\~nas, M. Barranco and P. Schuck,
                \NPA{519} (1990) 73c.
%
\bibitem{Cen93b} M. Centelles, X. Vi\~nas, M. Barranco and P. Schuck,
                 \AP{221} (1993) 165.
%
\bibitem{Spe92}  C. Speicher, R.M. Dreizler and E. Engel,
                 \AP{213} (1992) 312.
%
\bibitem{Von92b} D. Von-Eiff, S. Haddad and M.K. Weigel,
                 \PRC{46} (1992) 1797.
%
\bibitem{Cen92}  M. Centelles, X. Vi\~nas, M. Barranco, S. Marcos and
                 R.J. Lombard, \NPA{537} (1992) 486.
%
\bibitem{Von92a} D. Von-Eiff and M.K. Weigel, \PRC{46} (1992) 1797.
%
\bibitem{Cen93a} M. Centelles, X. Vi\~nas, M. Barranco, N. Ohtsuka,
                 Amand Faessler, Dao T. Khoa and H. M\"{u}ther,
                 \PRC{47} (1993) 1091.
%
\bibitem{Spe93}  C. Speicher, R.M. Dreizler and E. Engel, \NPA{562}
                 (1993) 569.
%
\bibitem{Cen93c} M. Centelles and X. Vi\~nas, \NPA{563} (1993) 173.
%
\bibitem{Lop88}  M. L\'{o}pez-Quelle, S. Marcos, R. Niembro, A.
                 Bouyssy and Nguyen Van Giai, \NPA{483} (1988) 479.
%
\bibitem{Von94b} D. Von-Eiff, J.M. Pearson, W. Stocker and
                 M.K. Weigel, \PLB{324} (1994) 279.
%
\bibitem{Mul93} H. M\"{u}ller and R.M. Dreizler,
                \NPA{563} (1993) 649.
%
\bibitem{Mul95} H. M\"uller and B.D. Serot, \PRC{52} (1995) 2072.
%
\bibitem{Eng94}  L. Engvik, E. Osnes, M. Hjorth-Jensen, G. Bao and
                 E. {\O}stgaard, Astrophys.\ J. 469 (1996) 794;
 H. Huber, F. Weber and M.K. Weigel, \NPA{596} (1996) 684;
 H. M\"uller and B.D. Serot, \NPA{606} (1996) 508.
%
\bibitem{Mat94}  F. Matera and V. Yu.\ Denisov, \PRC{49} (1994) 2816.
%
\bibitem{Bar81}  M. Barranco and J. Treiner, \NPA{351} (1981) 269.
%
\bibitem{Bon85}  P. Bonche, S. Levit and D. Vautherin, \NPA{436}
                 (1985) 265.
%
\bibitem{Sur87}  E. Suraud, \NPA{462} (1987) 109.
%
\bibitem{Lat78} J.M. Lattimer and D.G. Ravenhall,
                Ap.\ J.\ 223 (1978) 314.
%
\bibitem{Bar80} M. Barranco and J.R. Buchler, \PRC{22} (1980) 1729.
%
\bibitem{Das92} A. Das, R. Nayak and L. Satpathy, \JPG{18} (1992) 869.
%
\bibitem{Sat83} L. Satpathy and R. Nayak , \PRL{51} (1983) 1243.
%
\bibitem{Far86} M. Farine and J.M. Pearson, \PLB{167} (1986) 259.
%
\bibitem{Cal60} H.B. Callen, Thermodynamics (Wiley, New York, 1960).
%
\bibitem{Rav72} D.G. Ravenhall, C.D. Bennett and C.J. Pethick,
                \PRL{28} (1972) 978.
%
\bibitem{Sto73} W. Stocker, \NPA{215} (1973) 591.
%
\bibitem{Sto85} W. Stocker and M. Farine, \AP{159} (1985) 255.
%
\bibitem{Bar82}  J. Bartel, P. Quentin, M. Brack, C. Guet and H.-B.
                 H{\aa}kansson, \NPA{386} (1982) 79.
%
\bibitem{Bei75}  M. Beiner, H. Flocard, Nguyen Van Giai and
                 P. Quentin, \NPA{238} (1975) 29.
%
\bibitem{Lee86} S.J. Lee, J. Fink, A.B. Balantekin, M.R. Strayer,
                A.S. Umar, P.-G. Reinhard, J.A. Maruhn and W. Greiner,
                \PRL{57} (1986) 2916.
%
\bibitem{Sha93} M.M. Sharma, M.A. Nagarajan and P. Ring, \PLB{312}
                (1993) 377.
%
\bibitem{Jam89} M. Jaminon and C. Mahaux, \PRC{40} (1989) 354.
%
\bibitem{Ce90a} M. Centelles, M. Pi, X. Vi\~nas, F. Garcias and
                M. Barranco, \NPA{510} (1990) 397.
%
\bibitem{Lat81} J.M. Lattimer,
                Ann.\ Rev.\ Nucl.\ Part.\ Sci.\ 31 (1981) 337.
%
\bibitem{Gui95} A. Guirao, X. Vi\~nas and M. Pi, \ZPD{35} (1995) 199.
%
\end{thebibliography}
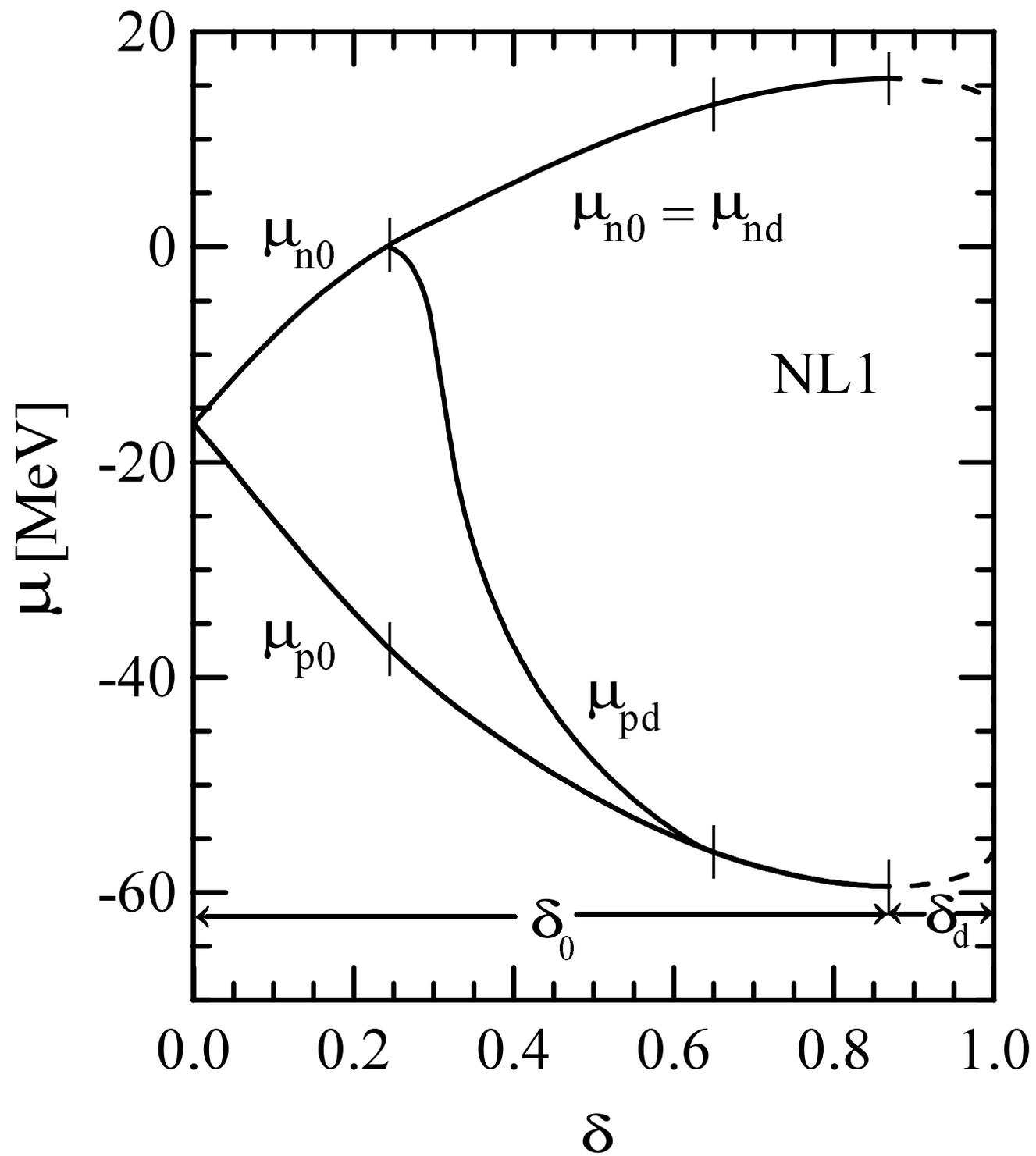
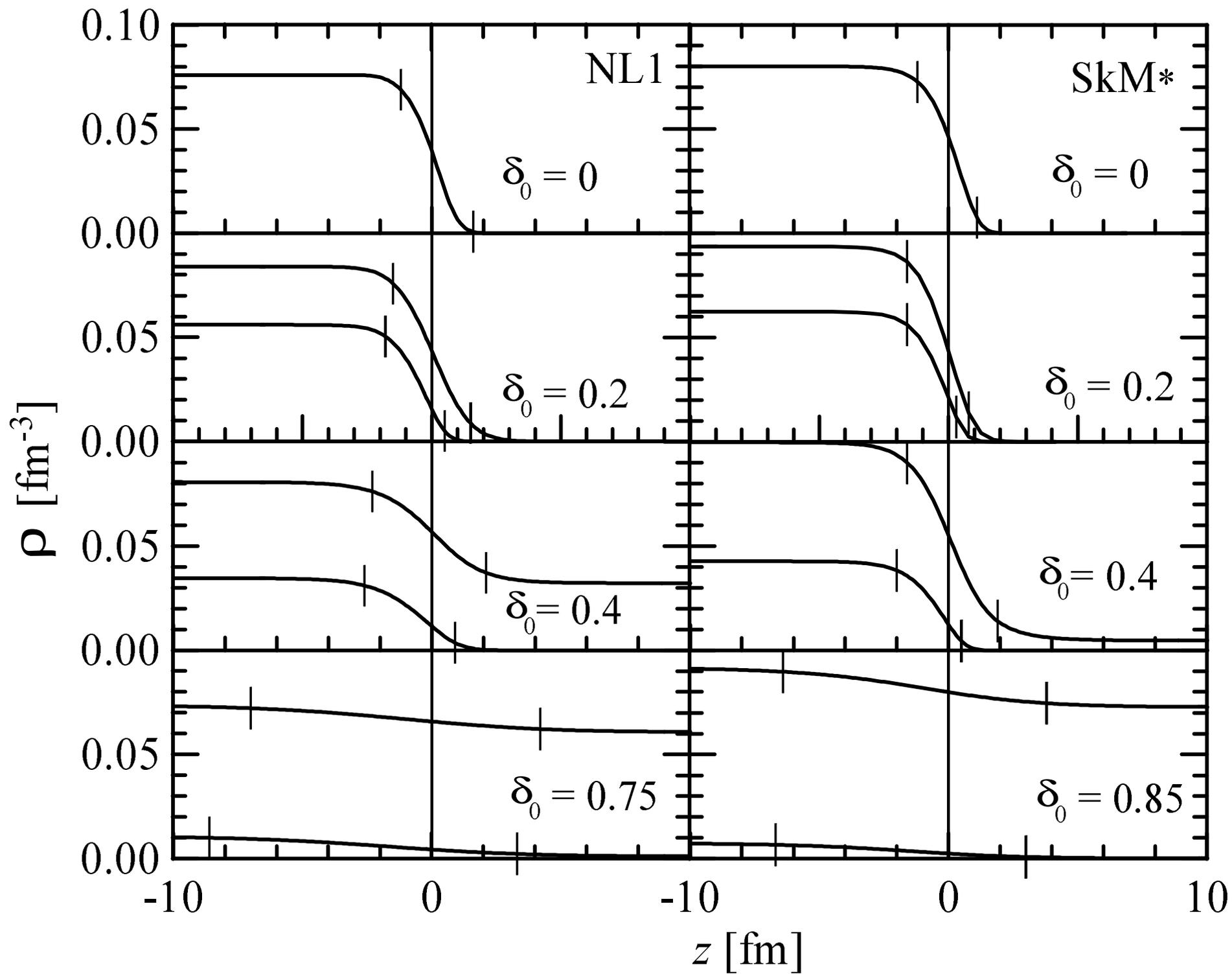
\end{document}